\documentclass[a4paper,11pt]{article}
\pdfoutput=1 

\usepackage{jcappub} 
\usepackage[T1]{fontenc}

\usepackage{cancel}
\usepackage{enumitem}    
\usepackage{amsthm}         
\usepackage{amsmath} 	   
\usepackage{amssymb}       
\usepackage{amsfonts}
\usepackage{graphicx} 	    
\usepackage{verbatim}
\usepackage{epsfig,multicol,bbm}
\usepackage{color}
\usepackage{colortbl}
\usepackage{float}
\usepackage{multirow}

\usepackage{subfig}

\allowdisplaybreaks

\title{Space-time slicing in Horndeski theories and its implications for non-singular bouncing solutions}

\author[a]{Anna Ijjas}

\affiliation[a]{Center for Theoretical Physics, Columbia University\\New York, NY, 10027, USA}

\emailAdd{Anna.Ijjas@Columbia.edu}

\abstract{
In this paper, we show how the proper choice of gauge is critical in analyzing the stability of non-singular cosmological bounce solutions based on Horndeski theories.  We show that it is possible to construct non-singular cosmological bounce solutions with classically stable behavior for all modes with wavelengths above the Planck scale where:  
(a) the solution involves a stage of null-energy condition violation during which gravity is described by a modification of Einstein's general relativity; and (b) the solution reduces to Einstein gravity both before and after the null-energy condition violating stage. Similar considerations apply to galilean genesis scenarios.
}

\begin{document}
\maketitle 
\flushbottom

\section{Introduction}
\label{sec:intro}
In recent years, Horndeski theories have received much attention due to their many cosmological applications, such as describing late-time acceleration \cite{Nicolis:2008in}, starting the universe from Minkowski space (galilean genesis) \cite{Creminelli:2006xe,Kobayashi:2015gga}, or admitting cosmological bounce \cite{Qiu:2011cy,Easson:2011zy,Koehn:2013upa} and wormhole solutions \cite{Rubakov:2015gza}.
Furthermore, it appears that the underlying Lagrangians have a natural connection to higher-dimensional theories with branes;  for example, a specific Horndeski theory, the conformal, flat-space galileon,  was found to describe scalar fluctuations on a static 3-brane embedded into an AdS$_5$ spacetime \cite{deRham:2010eu}.
A characteristic feature of these theories useful in many applications is that they introduce non-trivial couplings between scalar fields and the metric (`braiding') \cite{Deffayet:2010qz} while keeping the equations of motion second order. 

The goal of this paper is to identify a proper procedure for choosing the space-time slicing and fixing gauges so as to avoid introducing artificial coordinate singularities that lead to misinterpretation about the evolution of perturbations and linear stability in Horndeski theories. In particular, we will show that the most prevalent gauge conditions used in analyzing inflationary models (unitary and spatially-flat gauges) are poorly behaved in models where the Horndeski  theory asymptotes to Einstein gravity {\it both} in the far past and far future.  This includes the interesting case of geodesically complete bouncing and genesis models with a single scalar field. 
Notably, the transition from Einstein to Horndeski gravity includes a critical point (`$\gamma$-crossing') where the common gauge choices suggest a blow-up of long-wavelength modes even though there is none.  We will explicitly show that the blow-up is a mere coordinate singularity by identifying the gauge choices that avoid the problem.  The general lesson is that all gauge choices are not equally valid in Horndeski theories. When there is a blow-up, one must always check if it is a true physical divergence or simply a coordinate singularity.  This is especially important in theories that violate the null-energy condition (NEC).
 
The NEC implies that,  for every null-vector $k^{\mu}$, the stress-energy tensor
$T_{\mu\nu}$ obeys the inequality
\begin{equation}
T_{\mu\nu}k^{\mu}k^{\nu} \geq 0 \,.
\end{equation}
For perfect fluids, the NEC states that the sum of energy density ($\rho_{\rm tot}$) and pressure ($p_{\rm tot}$) always remains non-negative.
In a Friedmann-Robertson-Walker (FRW) spacetime that is given by the metric 
\begin{equation}
\label{FRW-metric}
ds^2 = -dt^2+a(t)dx^idx_i\,,
\end{equation}
where $a(t)$ is the scale factor and $t$ is physical time, and in which any form of stress-energy behaves like a perfect fluid (to leading order), the NEC reduces to the condition that the Hubble parameter $H=\dot{a}/a$ always decreases, or equivalently, 
\begin{equation}
\rho_{\rm tot} + p_{\rm tot} \equiv - 2 \dot{H}  \geq 0 \,,
\end{equation}
here dot denotes differentiation with respect to physical time $t$. That means, in an expanding universe ($H>0$), the total energy density ($\sim H^2$) decreases with time while, in a contracting universe ($H<0$), the total energy density increases with time. Obviously, if the NEC is satisfied in a contracting universe, as the universe is becoming smaller, the total energy density approaches Planckian values such that the classical curvature invariants eventually blow up.

Note that, although the NEC is satisfied by various forms of stress-energy, such as ordinary matter and radiation, it is not an implication of Einstein gravity but an independent condition that is commonly assumed for the purposes of simplicity or for proving certain theorems.  There is nothing to say the condition cannot be violated.  The situation is similar to the strong energy condition ($T_{\mu\nu}k^{\mu}k^{\nu} > 0$ or $\dot{H} > 0$) that is commonly assumed but which is known to be violated by the de Sitter solution, which has ($\dot{H} = 0$).
In particular, it is common to cite the Hawking-Penrose singularity theorems as no-go arguments against non-singular cosmologies. However, these theorems are closely tied to the NEC. That means, one way to avoid the conclusion of these theorems and obtain a well-behaved non-singular cosmology is to show that the NEC can be violated without introducing pathologies, such as Ostrogadski or quantum ghosts and classical gradient instabilities. 

Indeed, the possibility of stable violation of the NEC in Horndeski theories has been studied in the context of galilean genesis and cosmological bounce scenarios. 
More concretely, in the case of a contracting universe one must show that it is possible to transit to an expanding universe during a phase of NEC violation without bad behavior such as the blow-up of curvature invariants. Here, the transition point is called the `cosmological bounce' and the phase of NEC violation the `bounce stage.' The bounce is non-singular because it occurs at finite values of the scale factor, well below Planckian length or energy scales.

Whether pathologies can be truly avoided by these Horndeski cosmological solutions has been an open question.  For example, it has been shown by Libanov et al. \cite{Libanov:2016kfc} and later, in greater generality by Kobayashi \cite{Kobayashi:2016xpl} that violating the NEC in genesis and Horndeski bouncing  scenarios leads to a gradient instability or a singularity. We recently pointed out that, in Horndeski bouncing  scenarios, the bad behavior is not due to violation of the NEC \cite{Ijjas:2016tpn}, as was previously conjectured. 
Most importantly, no pathology occurs in bouncing scenarios where the braiding is turned off only after (or only before) the stage of NEC violation, {\it i.e.}, where the NEC violating cosmological solutions describe Einstein gravity only in the asymptotic future (or only in the asymptotic past). 
A similar result was found by Pirtskhalava et al. \cite{Pirtskhalava:2014esa} for galilean genesis.

Furthermore, in \cite{Ijjas:2016vtq} we showed that, to construct a geodesically complete bouncing cosmology that connects to Einstein gravity {\it both} before and after NEC violation, there must be a $\gamma$-crossing point ($\gamma=0$).  The quantity $\gamma$ characterizes the ratio between the three-curvature and braiding.  It is only at this $\gamma$-crossing point where there is potentially bad behavior.  Notably, $\gamma$-crossing need not occur during the NEC-violating phase (`bounce stage') but can be pushed arbitrarily far before (or after) the bounce stage. We shall see that this ability to separate $\gamma$-crossing from the bounce will play a critical role in the analysis below. 
 Recently, different groups argued that the pathology at $\gamma$-crossing can only be avoided in `beyond Horndeski' theories \cite{Cai:2016thi,Creminelli:2016zwa,Kolevatov:2017voe}.

In this paper, we study Horndeski theories admitting NEC violating solutions that reduce to Einstein gravity both before and after the NEC violating stage. We show that the stability analysis  around the $\gamma$-crossing point based on conventional unitary or spatially-flat gauge choices is misleading due to a coordinate singularity and prescribe the method to fix the problem. We will then show it is possible to construct stable classical cosmological bounce solutions that asymptote to Einstein gravity before and after the bounce without encountering any  pathologies for wavelengths greater than the Planck scale.  

The paper is organized as follows: In Sec.~\ref{sec:prelims}, we start with reviewing the basic features of Horndeski theories and derive the linearized field equations without fixing the gauge. We show next, in Sec.~\ref{sec:3+1}, that the unitary and spatially-flat gauges in which the no-go arguments were derived become ill-defined around the $\gamma$-crossing point and, hence, the claimed blow-up is a mere slicing issue.
Linearizing the theory in Newtonian gauge, which is well-behaved around $\gamma$-crossing, we identify a `transition scale.' Above and below this scale, scalar perturbations are governed by different evolution equations. The main result of this paper is to solve these equations around $\gamma$-crossing and to identify the scale ($k_T/a$) below which linear perturbations are well-behaved; we do this in Sec.~\ref{sec:newtonian}. We summarize our findings in Sec.~\ref{sec:summary}. 

\tableofcontents

\section{Preliminiaries}
\label{sec:prelims}

Horndeski theories are Lorentz-invariant gravitational theories with the characteristic feature that the corresponding Einstein equations as well as the scalar-field equations of motion all remain second-order in derivatives. In particular, Horndeski theories do not propagate Ostrogadski ghost degrees of freedom.   

The most general single-field Horndeski action is given by 
 \begin{equation}
 \label{H-action}
{\cal S} = \int d^4x  \sqrt{-g} \left(\frac12 R + {\cal L}_H + {\cal L} _{\rm matter}
\right)\,,
\end{equation}
where $g$ is the metric determinant; $R$ the Ricci scalar; ${\cal L} _{\rm matter}$ the Lagrangian that describes ordinary matter and radiation; and the Horndeski Lagrangian, 
\begin{equation}
{\cal L}_H = \sum_{i=2}^5 c_i {\cal L}_i 
\,,
\end{equation}
is defined as a linear combination of the Lagrangian densities 
\begin{eqnarray}
{\cal L}_2 &=& G_2(X,\phi)\,,\\
\label{L3-0}
{\cal L}_3 &=& G_3(X,\phi)\Box\phi\,,\\
\label{L4-0}
{\cal L}_4 &=& \frac12 G_4(X,\phi)R + G_{4,X}(X,\phi) \left( (\Box\phi)^2-(\nabla_{\mu}\nabla_{\nu}\phi)^2\right) \,,\\
\label{L5-0}
{\cal L}_5 &=& G_5(X, \phi)G_{\mu\nu}\nabla^{\mu}\nabla^{\nu}\phi -\frac{1}{3}G_{5,X}(X,\phi) \left( (\Box\phi)^3 - 3 \Box\phi (\nabla_{\mu}\nabla_{\nu}\phi)^2+2(\nabla_{\mu}\nabla_{\nu}\phi)^3 \right)
\,.\quad
\end{eqnarray}
Here, each $G_i(X, \phi)$ characterizes the $i$th Horndeski interaction; 
$
X=-(1/2) \nabla_{\mu}\phi\nabla^{\nu}\phi 
$
is the canonical kinetic term; $G_{\mu\nu}$ is the Einstein tensor; and the $c_i$ ($i=2,...,5$) are real constants. Throughout, we use the mostly positive metric signature $(- +++)$ and we work in reduced Planck units $M_{\rm Pl}^2=8\pi G_{\rm N}=1$, where $G_{\rm N}$ is Newton's constant.
For reasons that we will clarify below, we separated the Einstein-Hilbert part, in particular, the action of a free canonical scalar is recovered by setting $G_2\equiv X, G_3 \equiv G_4\equiv G_5 \equiv 0$. In addition, we assumed that matter is universally coupled, {\it i.e.}, its action is independent of the scalar $\phi$. 

We note that Horndeski theories are sometimes also called `generalized galileons.' 
The reason is that Horndeski theories were independently re-discovered based on symmetry arguments, a few decades after they were first identified by Horndeski as the most general Lorentz invariant theories with second-order equations of motion \cite{Horndeski:1974wa}. To avoid confusion, we omit  the use of the term `generalized galileons' as a synonym for Horndeski theories. Rather, by galileons, we mean only special Horndeski Lagrangians that are invariant under the `galilean' transformation 
\begin{equation}
\label{galileon-sym}
\phi \to \phi + b^{\mu} x_{\mu} + c \,,
\end{equation}
where $b_{\mu}$ $(\mu = 0, ..., 3)$ and $c$ are real constants.

Intriguingly, galileons appear to have a natural connection to higher-dimensional theories involving branes. In Ref.~\cite{deRham:2010eu}, it was found that  fluctuations of static three-branes probing five-dimensional space-times can be described by simple scalar-field Lagrangians that  
correspond to particular galileons, where the specific interactions $G_i$ are fixed by the chosen 5d background metric and the brane metric.
Different choices of these metrics lead to different types of galileons; for a review see~Ref.~\cite{Goon:2011uw}. It remains, though, an open question whether all Horndeski theories can be derived in the same or in a similar way such that they could be understood as a particular manifestation of some three-brane geometry.

More general Horndeski Lagrangians have the same structure as the conformally-coupled ${\cal L}_4$ Horndeski action up to linear order.  The differences due to disformal couplings in ${\cal L}_4$ or ${\cal L}_5$ theories appear first at third-order in perturbation theory \cite{Gao:2012ib} that is not within the scope of the present work.  Furthermore, it has recently been shown that the conformally-coupled ${\cal L}_4$ Horndeski theory ($\partial_X G_4 \equiv 0, G_5 \equiv 0 $) is linearly well-posed \cite{Papallo:2017qvl}.  This feature makes the conformally-coupled ${\cal L}_4$ Horndeski theory particularly attractive for numerical applications.   More general Horndeski theories (with  $\partial_X G_4 \neq 0$ or $G_5 \neq 0 $), on the other hand, appear to be  ill-posed. 

For cosmological applications that are the focus of this paper -- classical theories that admit non-singular bouncing or genesis solutions that reduce to Einstein gravity both before and after the bounce or genesis phase -- the richest structure needed is exactly the conformally-coupled ${\cal L}_4$ Horndeski action \cite{Ijjas:2016vtq}.  
Hence, in the remainder of this paper, we study the particular Horndeski theory, specified through the couplings
\begin{equation}
\label{G-i}
G_2(X,\phi),\quad G_3(X,\phi) \equiv -b(\phi)X, \quad G_4(X,\phi)\equiv G_4(\phi), \quad \rm{and}  \quad G_5(X,\phi)\equiv 0 \,.
\end{equation}

\subsection{Covariant dynamics}

Varying the action~\eqref{H-action} with respect to the metric $g^{\mu\nu}$ yields the Einstein field equations 
\begin{equation}
G_{\mu\nu} = T_{\mu\nu} \equiv - 2\frac{\delta \left({\cal L}_H +  {\cal L}_{\rm matter}\right)}{\delta g^{\mu\nu}} + \left({\cal L}_H +  {\cal L}_{\rm matter}\right)g_{\mu\nu}
\end{equation}
with the stress-energy tensor taking the form
\begin{eqnarray}
\label{cov-T_munu}
T_{ \mu \nu}  & = & \Big(G_2(X,\phi) + b(\phi)  \nabla _{\mu} \phi \nabla ^{\mu} X - 2 b_{, \phi}(\phi) X ^2 + 2G_{4,\phi\phi}(\phi)X - G_{4,\phi}\Box\phi \Big) g _{\mu\nu} 
\nonumber\\
&+& \Big( G_{2,X}(X,\phi) -b( \phi) \Box \phi - 2 b_{,\phi}(\phi) X + G_{4,\phi\phi}(\phi) \Big)\nabla _{\mu} \phi\, \nabla _{\nu} \phi 
\nonumber\\
 &-&  b( \phi) \Big(\nabla _{ \mu} \phi\, \nabla _{ \nu} X +  \nabla _{ \nu} \phi\, \nabla _{ \mu} X\Big) + G_{4,\phi} \nabla _{\mu} \nabla_{\nu} \phi  - G_4(\phi) G_{\mu\nu} + T_{ \mu \nu}^{\rm (matter)}
 \,.
\end{eqnarray}
Generically, the stress-energy tensor of Horndeski theories does not take the form of a perfect fluid. That is, unlike in the case of canonical free scalars or so-called $P(X,\phi)$ theories, there is no reference frame in which the the Horndeski stress-energy tensor is diagonal.

The choice of the reference metric $g_{\mu\nu}$ is especially important for cosmological model building since the notions of contraction and expansion are not invariants but only make sense when defined with respect to a reference scale. There is a common misbelief that one must always analyze gravitational Lagrangians with a single scalar-field in the `Einstein frame,' {\it i.e.}, in the representation where the four-Ricci scalar is minimally coupled to the field, and so the proper reference metric is the Einstein-frame metric. In reality, it is the matter Lagrangian that unambiguously yields the `physical' reference scale.  Once the matter is specified in the action, it is unambiguous whether the universe is expanding or contracting:
The physical reference metric is the one matter, radiation, or any test-particle `sees,' the metric choice in which the corresponding matter Lagrangian is independent of the scalar field and vice-versa.  For example, a straightforward way to find the physical reference scale and the corresponding stress-energy tensor is to choose the representation where the matter is universally coupled, as was done above in Eq.~\eqref{H-action}. In modified gravity theories, this is often not the Einstein-frame.  
Remarkably, there are known examples where a conformal transformation of the metric changes the sign of the corresponding Hubble parameter such that simply specifying the matter coupling can lead to different cosmologies \cite{Ijjas:2015zma}.

Finally, variation of the action with respect to the scalar yields the evolution equation for $\phi$,
\begin{eqnarray}
-G_{2,X}  \Box\phi &=&    \left( G_{2,XX} - 2b_{,\phi} \right) \nabla_{\mu}X\nabla^{\mu}\phi 
- 2X\left( G_{2,X\phi} -b_{,\phi\phi}X \right) + G_{2,\phi}  + \frac12 G_{4,\phi} R 
\nonumber\\
& -& b(\phi) \left( (\Box \phi)^2 - (\nabla_{\mu}\nabla_{\nu}\phi)^2 - R_{\mu\nu}\nabla^{\mu}\phi\nabla^{\nu}\phi \right)
\,.
\end{eqnarray}

\subsection{Background evolution}
Evaluating the Einstein equations for a homogeneous FRW metric as given in Eq.~\eqref{FRW-metric} yields the Friedmann equations 
\begin{eqnarray}
\label{FRW1}
3 H^2  &=&  - G_2(X,\phi) + G_{2,X}(X,\phi) \dot{\phi}^2 - \frac{1}{2} b_{,\phi}(\phi) \dot{\phi}^4 + 3 H b(\phi) \dot{\phi}^3 \\
&-& 3G_{4,\phi}H\dot{\phi} - 3G_4(\phi)H^2 + \rho_{\rm matter}
\nonumber
\,, \\
\label{FRW2}
-2 \dot{H}   &=&  G_{2,X}(X,\phi) \dot{\phi}^2 - b_{,\phi}(\phi) \dot{\phi}^4 + 3Hb(\phi)\dot{\phi}^3- G_{4,\phi}\dot{\phi}H + G_{4,\phi\phi}\dot{\phi}^2
\\
&+& \left(G_{4,\phi} - b(\phi)\dot{\phi}^2 \right)\ddot{\phi} + 2 G_4(\phi)\dot{H} + \left(\rho_{\rm matter} + p_{\rm matter}\right)
\,,\nonumber
\end{eqnarray}
that, in the case of a single scalar, are sufficient to fully determine the background dynamics.
The right hand side of the first Friedmann equation describes the relative evolution of the different components that make up the total total energy density ($3H^2 = \rho_{\rm tot}$). The second Friedmann equation describes the evolution of total energy density that is given by the sum of total energy density and pressure ($\dot{H}= \rho_{\rm tot}+ p_{\rm tot}$).

The evolution equation for the homogeneous scalar $\phi(t)$ reads as
\begin{eqnarray}
\label{FRW3}
&& \left(G_{2,X} + \left( G_{2,XX} - 2 b_{,\phi}\right) \dot{\phi}^2 + 6 bH \dot{\phi}   \right) \ddot{\phi}+   3H\dot{\phi}  G_{2,X}  
+  3 b(\phi) \dot{\phi}^2 \left( \dot{H} + 3H^2 \right) =   
\nonumber\\ 
 &=& G_{2,\phi} - \left( G_{2,X\phi} - \frac{1}{2} b_{,\phi\phi} \dot{\phi}^2\right) \dot{\phi}^2 +3 G_{4,\phi}  \left(\dot{H} + 2H^2 \right)
\,;
\end{eqnarray}
note that Eq.~\eqref{FRW3harm-0} is not independent but can be derived using the two Friedmann equations. 

Although the stress-energy tensor of Horndeski theories does not take the form of a perfect fluid, the homogeneous part does because it inherits it from the spherically symmetric FRW background geometry. Deviations from a perfect fluid (given an FRW background) only become evident when one considers spatial inhomogeneities.

Typically, as a first step of model building, one identifies those Lagrangians that admit the desired cosmological background solutions. In particular, this means checking that a certain FRW solution exists for a given choice of the couplings $G_2(X, \phi), b(\phi)$, and $G_4(\phi)$.
For a non-singular bounce solution, for example, this means picking a Lagrangian that transits a contracting FRW universe ($H<0$) to an expanding FRW universe ($H>0)$. 

Of course, as we have noted before, the goal is to have the bounce occur at energies well below the Planck scale.

\subsection{Linearized field equations}

If a Lagrangian admits a particular background solution, the next logical step is to check whether the solution is linearly stable. This means, the modes have second-order equations of motion (no Ostrogadski instability) and they suffer from no classical gradient instability or, equivalently, the modes propagate at non-negative sound speed. In addition, one typically requires strict positivity for the kinetic coefficient of linear perturbations to avoid quantum ghost instabilities.
If the background solution fails the linear stability test, generically, trajectories will not follow the homogeneous solution.      


To study linearized metric and matter perturbations, we use the $3+1$~decomposition or ADM~slicing of the metric. 
Here, spacetime is foliated by spatial hypersurfaces $\Sigma_t$ of constant time coordinate $t$ while the constant-time hypersurfaces are threaded by lines of constant spatial coordinates $x^i$ such that the line element is given by
\begin{eqnarray}
ds^2 &=& -N^2 dt^2 + \gamma_{ij}\big( dx^i + N^idt \big)\big(dx^j + N^jdt \big) 
\,,
\end{eqnarray}
where $N$ is the lapse function and measures  the ratio between the proper time and coordinate time along the future directed, timelike unit normal vector $n_{\mu}$ to the spatial hypersurface $\Sigma_t$; the shift vector $N^i$ gives the difference between proper and coordinate time; and $\gamma_{ij}$ is the spatial metric of the constant time hypersurface. Note that $n_{\mu}$ need not coincide with the time vector tangent to the threading lines. The freedom to choose the lapse $N$ ({\it i.e.}, the foliation $\Sigma_t$) and the shift $N^i$ ({\it i.e.}, the spatial coordinates $x^i$) reflects the coordinate freedom of general relativity.

Linearizing the metric and decomposing the perturbations into scalar, vector, and tensor modes, the line element takes the form
\begin{eqnarray}
\label{adm-lin-met}
ds^2 &=& -\big(1+2\alpha\big)dt^2 + 2a(t)\big( \partial_i \beta + S_i \big) dtdx^i \\
&+& a^2(t)\left( \big(1 - 2\psi \big)\delta_{ij} + 2\partial_i\partial_j\varepsilon + 2F_{(i,j)} + 2h_{ij} \right) dx^idx^j
\,,\nonumber
\end{eqnarray}
where $\delta N \equiv \alpha$ is the linearized lapse perturbation; $\delta N_i \equiv \partial_i \beta + S_i $ is the linearized shift perturbation; and $\delta\gamma_{ij} \equiv  -\psi \delta_{ij} + \partial_i\partial_j\varepsilon + F_{(i,j)} + h_{ij} $ is the linearized perturbation of the spatial metric $\gamma_{i j}$. The four perturbation variables $\alpha, \beta, \psi,$ and $\epsilon$ represent scalar quantities; the two perturbation variables $S_i$ and $F_i$ (with $S^i_{,i}=F^i_{,i}=0$) represent vector quantities; and $h_{ij}$ (with $h^i_i=h^i_j{}_{,i}=0$) is the transverse, traceless tensor perturbation. Notably, in this decomposition,  scalar, vector, and tensor perturbations evolve independently at linear order and hence we can study them separately. 
We denote the linearized perturbation of the scalar field $\phi$ by $\pi(x,t)$. 

In the following, we will not consider vector perturbations since we assume that those are suppressed by some smoothing mechanism, e.g., in the case of the bounce, by the preceding contracting phase, and, in the case of genesis, by the super-acceleration. For the same reason, we will not consider perturbations of ordinary, pressure-less matter and radiation. They do not directly interact with the Horndeski scalar and, hence, remain linearly stable throughout. 
We also note that in conformally-coupled ${\cal L}_4$ Horndeski theories where the effective Planck mass, ({\it i.e.}, the coupling to the four-Ricci scalar in the action) is greater than zero, tensor perturbations evolve as in Einstein gravity, stably propagating at a sound speed $c_T=1$; see, {\it e.g.}, \cite{Kobayashi:2011nu}.

The scalar part of the linearized Einstein equations \eqref{cov-T_munu} take the following form:
\begin{eqnarray} 
\label{0-0-u-simple}
& & \Big( 6 H\gamma(t) - 3A _h(t)H^2(t) - \rho_K(t)  \Big) \alpha + \gamma(t)\left(3\dot{\psi} + \frac{k^2}{a^2}\sigma \right) +  \frac{k^2}{a^2} A_h(t) \psi  
\qquad\\
&+&  \Big( 3H(t)\gamma(t) -3A_h(t)H^2(t)  - \rho_K(t) \Big)  \delta\dot{u} 
 -   \left( 3\dot{H}(t) \gamma(t) + \frac{ k^2}{a^2}\Big(A_h(t)H(t) - \gamma(t) \Big)  \right) \delta u 
= 0
\,,\nonumber\\
\nonumber\\
\label{t-i-u-simple} 
& &  A_h(t) \dot{\psi} 
-  \Big( A_h(t)H(t) - \gamma(t)  \Big) \delta \dot{u} 
= -\gamma(t)\alpha + A_h(t) \dot{H}(t)  \delta u 
\,,\\
\nonumber\\
\label{offdiag-u-simple}
&& A_h(t)  (\alpha - \psi - \dot{\sigma} - H\sigma)  =  \dot{A}_h(t) \left(\delta u + \sigma\right)
\,, 
\\
\nonumber\\
\label{diag-u-simple0}
& & \big(\gamma(t)\alpha\big)^{\cdot} +   3H \big(\gamma(t) \alpha \big) +
A_h(t)\ddot{\psi}  +  \left(3A_h(t)H + \dot{A}_h(t) \right) \dot{\psi} 
\\
&-&  \left( A_h(t)H(t) - \gamma(t)  \right)  \delta\ddot{u}  
+  \left(  \dot{\gamma}(t) - 3H\left( A_h(t)H(t) - \gamma(t)  \right) - 2A_h(t)\dot{H}(t) - \dot{A}_h(t)H(t)
 \right) \delta\dot{u} 
\nonumber\\
&-&  \left( A_h\ddot{H} + \dot{A}_h\dot{H}   + 3A_hH\dot{H} \right)\delta u
 = 0 \,.
 \nonumber
\end{eqnarray}
where Eq.~\eqref{0-0-u-simple} is the linearized Hamiltonian constraint; Eq.~\eqref{t-i-u-simple} is the linearized momentum constraint; Eq.~\eqref{offdiag-u-simple} is the linearized anisotropy equation; and Eq.~\eqref{diag-u-simple0} is the linearized pressure equation; $\delta u  \equiv - \delta \phi/\dot{\phi} $; and the scalar shear perturbation $\sigma$ is defined through
\begin{equation}
\label{shear}
\frac{\sigma(t,{\bf x})}{a(t)}\equiv  a(t)\dot{\epsilon}(t,{\bf x}) - \beta(t,{\bf x}) \,;
\end{equation} 
$\sigma$ is the scalar component of the linearized shear tensor
\begin{equation}
\sigma_{\mu\nu} = \frac13 K\gamma_{\mu\nu} - K_{\mu\nu}
\,.
\end{equation}
The quantities
\begin{eqnarray}
\label{A_h-def}
A_h(t) &=& 1+G_4(\phi)
\,,\\
\label{gamma}
\gamma(t) &=& A_h(t) H(t) - \frac12\left( b( \phi)\dot{\phi}^3(t) - \dot{A}_h(t) \right) 
\,,\\
\label{rho_K}
\rho_K(t) &=&  \frac12G_{2,X} \dot{\phi}^2 + \frac12 \big(G_{2,X X} - 2b_{,\phi}  \big) \dot{\phi}^4  + 3Hb(\phi)\dot{\phi}^3
\end{eqnarray}
are functions of the homogeneous background solution. We will see in the next section that the quantity $\gamma$ plays a crucial role in theories admitting NEC-violating solutions that reduce to Einstein gravity {\it both} before and after NEC violation because the transition from Einstein to Horndeski gravity implies that $\gamma$ has to go through zero (`$\gamma$-crossing').  We show that these solutions involve a coordinate singularity at the $\gamma$-crossing point in all but one algebraic gauge. This coordinate singularity has been mistaken as a real, physical blow up and led to wrong conclusions about the stability of the corresponding cosmological solutions.

The linearized scalar-field equation is given by 
\begin{eqnarray}
\label{scalar-eq-u0}
&&  \left( \rho_K + 3H \big(A_hH - \gamma\big) \right) \dot{\alpha} -\big( A_hH - \gamma \big)\frac{k^2}{a^2}\alpha
\\
&+& \left( \dot{\rho}_K + 3H\rho_K + ( 6\dot{H} + 9H^2) \big(A_hH - \gamma \big)  
+  3H \big( \dot{A}_hH -\dot{\gamma} \big) \right) \alpha
\nonumber\\
& + & \Big(A_h(t)H(t) - \gamma(t)\Big) \frac{k^2}{a^2}(\dot{\sigma} + H(t)\sigma) + \frac{k^2}{a^2}\left( \dot{A}_hH -\dot{\gamma}\right)\sigma
\nonumber\\
& + &  3\big(A_h(t)H(t) - \gamma(t)\big) \ddot{\psi}   - \dot{A}_h(t) \frac{k^2}{a^2}  \psi 
+ 3 \left(  3H(t) \big(A_h(t)H(t) - \gamma(t)\big) + \dot{A}(t)H(t) -\dot{\gamma}(t) \right) \dot{\psi}
\nonumber\\
&+&   \rho_K(t) \delta\ddot{u} + \left( \dot{\rho}_K(t)  + 3H \rho_K(t)  \right) \delta\dot{u} 
+  \Big( H\big(A_h(t)H(t) - \gamma(t)\big) + 2\dot{A}_hH  -\dot{\gamma}  \Big)  \frac{k^2}{a^2}\delta u
\nonumber\\
&-&  3\left( \big(\dot{A}_hH - \dot{\gamma}\big)\dot{H} + \big(A_hH - \gamma\big) \big(\ddot{H} + 3 H\dot{H}\big)\right)  \delta{u}  
= 0
\,.
\nonumber
\end{eqnarray}
For a detailed derivation of the linearized equations~(\ref{0-0-u-simple}-\ref{diag-u-simple0}) and \eqref{scalar-eq-u0} see Appendix~\ref{app-cov-Eqs}.

Note that there remain two residual scalar degrees of freedom related to the fact that the gauge has not yet been fixed.  
The subtleties of how to fix the gauge in Horndeski theories is the subject of the next section.

\section{Gauge choice(s) in Horndeski theories}
\label{sec:3+1}

In applications of Horndeski theories to the early-universe with a single scalar field, linear perturbations are most commonly analyzed in the {\it unitary gauge} (defined by $\delta u = 0$) where all spatial inhomogeneities are promoted to the metric while the scalar field remains unperturbed. Since this gauge `freezes out' perturbations of the scalar field, the otherwise involved computation of perturbed higher-derivative operators turns remarkably simple. 
At the same time, as we will discuss below, the lapse and shift perturbations act like Lagrange multipliers and can be eliminated using the constraint equations such that the only dynamical variable is the gauge-invariant quantity $\zeta \equiv - \psi$. We note that the unitary gauge is also utilized in low-energy effective field theory with the `Stueckelberg trick' being used to recover gauge invariance; see, {\it e.g.}, \cite{Piazza:2013coa}.

 Another preferred choice to study Horndeski early-universe scenarios is the {\it spatially-flat gauge} (defined through $\psi \equiv 0$) where all spatial inhomogeneities are promoted to the scalar field while the spatial metric remains unperturbed. Similarly to the unitary gauge, the lapse and shift perturbations are Lagrange multipliers in spatially-flat gauge and can be eliminated such that the only dynamical degree is $\delta \phi$. This gauge is particularly well-suited to study models with with multiple fluctuating scalar fields.
 
In certain physical situations, though, one has to be more careful because gauge-invariant quantities can become ill-defined.
An analogous problem occurs in Einstein gravity with a single scalar field during reheating when the field starts to oscillate at the bottom of the potential. At the turning points, the kinetic energy (or velocity) of the scalar $\dot{\phi}$ passes through zero and the spatial hypersurface of constant co-moving velocity ($\delta u \equiv \delta\phi/\dot{\phi} =0$) ceases to be space-like. 
Formally, the co-moving curvature mode ${\cal R}$, which corresponds to local perturbations of the scalar factor, appears to blow up. We know, though, that the blow-up is not physical.  Rather, it is a coordinate singularity resulting from a slicing issue. This fact becomes evident if one chooses a different gauge in which the spatial hypersurface remains space-like at the turning points and finding that all scalar model are under perturbative control.  
 
 In this section, we will identify a similar, turning-point-like problem in Horndeski theories and demonstrate that an ill-chosen gauge can lead to wrong conclusions about the linear stability behavior in the vicinity of the `turning point' (associated with $\gamma$-crossing). The main result of this paper is to perform the calculation in a gauge that is well-defined at and around the `turning-point' and to show that the linear stability behavior is different than previously thought.   

We start with a systematic analysis of the commonly used gauges in Horndeski theories.
A gauge describes a correspondence between the coordinates of the physical spacetime and a given background spacetime (described here by a FRW metric) \cite{bardeen:1980kt}. Fixing the gauge means picking a particular foliation for the physical spacetime, {\it i.e.}, setting the time slicing (lapse)  and the spatial coordinates (shift). 
Following Ref.~\cite{Khokhlov:2001ws}, we distinguish two ways of gauge fixing: In the first case, two additional constraint equations are introduced such that both the lapse and the shift remain non-dynamical and can be expressed as functions of the coordinates and local values of the perturbed scalar variables $\psi, \epsilon, \pi$ and their derivatives.  These are known as {\it algebraic gauges}. In the second case, two additional partial differential equations are introduced that make both the lapse and the shift dynamical. These are known as {\it differential gauges}. 

The main result of this section will be to show that NEC violating solutions that reduce to Einstein gravity both before and after the stage of NEC violation uniquely pick an algebraic gauge, namely the Newtonian gauge, while all other algebraic gauges commonly preferred in the literature involve a coordinate singularity. 

\subsection{Conformally-coupled ${\cal L}_4$ Horndeski in algebraic gauges}

In algebraic gauges, the system of linearized Einstein equations is supplemented by two additional, non-dynamical constraint equations. This has the advantage that it is fairly straightforward to derive the evolution equation for the single propagating degree of freedom (represented by one of the scalar gauge variables) and express the other gauge variables as a function of this variable.

In practice, one can, for example, use the linearized Hamiltonian and momentum constraints to eliminate the lapse perturbation $\alpha$ and the gradient of the scalar shear perturbation $\sigma$, by expressing both as a function of $\delta u, \delta \dot{u}, \psi$ and $\dot{\psi}$:
\begin{eqnarray}
\label{lapse-eq}
 \alpha &=& \frac{A_h(t)}{\gamma(t)} \left( -\dot{\psi}  + H(t)  \delta \dot{u} + \dot{H}(t)  \delta u \right) - \delta \dot{u} 
 \,,\\
\label{shear-eq}
\frac{k^2}{a^2}\sigma  & = & 
 \frac{r(t) }{\gamma^2(t)} \left( -\dot{\psi}  + H(t)  \delta \dot{u} + \dot{H}(t)  \delta u \right)
+  \frac{k^2}{a^2} \frac{A_h(t)}{\gamma(t)}\left(- \psi + H(t)\delta u \right) - \frac{ k^2}{a^2}\delta u\,,
\end{eqnarray}
where we define
\begin{equation}
\label{scriptA}
r(t) \equiv A_h(t)\rho_K(t) + 3\Big(A_h(t)H(t) -\gamma(t) \Big)^2
\,.
\end{equation}
These two equations encapsulate the main difference between Horndeski theories and  scalar-field theories with a stress-energy tensor that takes a perfect-fluid form. In particular, the equations determine the relation between the three quantities $A_h, \gamma$ and $\rho_K$ we introduced in Eqs.~(\ref{A_h-def}-\ref{rho_K}). For a perfect-fluid type scalar, there can be no mixing between the kinetic energy of the scalar and the metric. In particular, the function $\rho_K(t) \equiv r(t)$ that measures the kinetic energy of the field is independent of the metric and the function $\gamma(t)$ that measures the kinetic energy of the metric is field independent, $\gamma(t) = H(t)$.

In Horndeski theories, this is not anymore the case: both $\rho_K$ and $\gamma$ involve metric and scalar kinetic terms.  The direct mixing between the kinetic energy of the scalar and the metric -- a feature also called `braiding' \cite{Deffayet:2010qz} -- can be characterized by the deviation of $\gamma$ from $H$. 
In addition, the conformal quartic Horndeski interaction leads to a mixing between the scalar field and the four-Ricci scalar as meaured by the deviation between $A_h$ and unity.\footnote{
Note that our nomenclature deviates from the one introduced in Ref.~\cite{Bellini:2014fua} that is commonly used in the dark energy/modified gravity literature; the dictionary between the different notions is as follows: $M_{\ast}^2\ = 2A_h; HM_{\ast}^2\alpha_B = 2(A_hH-\gamma)$; and  $H^2M_{\ast}^2\alpha_K = 2\rho_K $. We changed the definition of the three background functions because it is more appropriate for the context of early-universe applications of Horndeski Lagrangians.
}

Substituting the expressions for $\alpha$ and $\sigma$ into the anisotropy equation, 
we obtain a simple second-order differential equation for the gauge-invariant quantity $-\psi + H\delta u$:
\begin{eqnarray}
&&  \frac{{\rm d}^2}{{\rm d}t^2}\Big( -\psi  + H(t)  \delta u  \Big)  
+\frac{d}{dt}\ln \left( a^3(t)A_h(t) \frac{r(t) }{\gamma^2(t)}\right) \frac{{\rm d}}{{\rm d}t}\Big( -\psi  + H(t)  \delta u  \Big)
\qquad\\
&+& \frac{2\dot{A}_h(t)\gamma(t) - A_h(t) \dot{\gamma}(t) - \big(A_h(t)H(t)-\gamma(t)\big)\gamma(t)}{r(t)} \frac{k^2}{a^2} \Big(- \psi + H(t)\delta u \Big)
 = 0\,.
\nonumber
\end{eqnarray}
Rescaling the variable 
\begin{equation}
-\psi + H\delta u \to v\equiv z (-\psi + H\delta u)\,,
\end{equation}
 where $z^2(t) = a^3(t)A_h(t) r(t) /\gamma^2(t)$, yields the Horndeski generalization of the Mukhanov-Sasaki variable with the corresponding wave equation
\begin{equation}
\label{v-equation}
\ddot{v}_k + \left(c_{\zeta}^2(t)\frac{k^2}{a^2} -\frac{\ddot{z}}{z}  \right)v_k = 0
\,.
\end{equation}
The sound speed of the modes  is given by 
\begin{equation}
\label{c-zeta}
c_{\zeta}^2(t) = \frac{2\dot{A}_h(t)\gamma(t) - A_h(t) \dot{\gamma}(t) - \big(A_h(t)H(t)-\gamma(t)\big)\gamma(t)}{r(t)}
\,.
\end{equation}

The simplicity of the Mukhanov-Sasaki equation is one of the reasons why the unitary ($\delta u = 0$) and spatially-flat ($\psi = 0$) gauges are the most common choices in studying linearized Horndeski theories: In either of the two gauges, solving for $v_k(t)$ immediately yields a solution for the dynamical gauge variable $\zeta = - \psi$ or $\delta u$, respectively.  Hence, in these two gauges, one can read off gradient instability from the sign of $c_{\zeta}^2$ as long as the slicing is valid. 

In other algebraic gauges, one needs the gauge condition to determine the dynamics of all gauge variables after solving for $v_k(t)$. 
Less commonly used algebraic gauges are 
\begin{itemize}
\item the {\it Newtonian gauge} that is defined through $\sigma = 0$; 
\item the {\it synchronous gauge} defined by $\alpha = 0$ that picks the spatial hypersurface of constant time; 
\item the {\it uniform density gauge} ($\delta \rho \equiv 0$); and 
\item the {\it uniform Hubble gauge} given by
$\delta H = H\alpha + \dot{\psi} + \frac13 \frac{k^2}{a^2}\sigma \equiv 0$.
\end{itemize}

We note that the algebraic gauge conditions discussed here only fix the time slicing but not the spatial coordinates. This is because the linearized Einstein equations and the gauge constraints together only determine the evolution of the scalar shift implicitly, by determining the shear perturbation $\sigma$. An additional constraint is needed that determines the spatial coordinates.  In practice, to fix the spatial coordinates, one usually chooses $\epsilon\equiv 0$.

 \subsection{$\gamma$-crossing}
 
 Even though the unitary and spatially-flat gauges are convenient choices, these gauges come with a serious drawback. It is immediately obvious from the constraint equations~(\ref{lapse-eq}-\ref{shear-eq}) that at and around the time $ t_{\gamma}$ when $\gamma$ hits zero  ($\gamma$-crossing), both gauge variables $\alpha$ and $\sigma$ blow up since, generically, the Mukhanov-Sasaki equation admits at least one non-zero solution. This meets the textbook definition of a coordinate singularity. A corollary is that, in the vicinity of $t = t_{\gamma}$, one cannot use the unitary and spatially-flat gauges to analyze the linear stability of the background solution.

In fact, there is only a single algebraic gauge that protects $\sigma$ from becoming singular at $\gamma$-crossing, namely the gauge where $\sigma \equiv 0$, {\it i.e.}, the Newtonian gauge. 
 In any other algebraic gauge, the constraint equation~\eqref{shear-eq} implies either a blow-up of $\sigma$ or a blow-up of $\delta u$ or $\psi$ or both around $t_{\gamma}$. This is because $\sigma$ depends quadratically on $\gamma^{-1}$ while $\alpha$ only linearly so setting, {\it e.g.}, $\alpha\equiv 0$ (synchronous gauge) is not sufficient.
 
 This is a crucial point since all no-go theorems stating that NEC violation always implies a classical gradient instability or a singularity of the cosmological solution were formulated either in unitary  \cite{Libanov:2016kfc,Kobayashi:2016xpl,Kolevatov:2016ppi} or in spatially-flat gauge  \cite{Akama:2017jsa}. 
At the same time, our finding recovers and explains the result of Ref.~\cite{Ijjas:2016tpn}, namely that any bad behavior that one encounters when studying linearized Horndeski theories in unitary or spatially-flat gauges is related to $\gamma$-crossing and not to NEC violation, as previously conjectured. As pointed out in Ref.~\cite{Ijjas:2016vtq}, these are the cosmological solutions that reduce to Einstein gravity both before and after the NEC violating stage.

Of course, to show that the pathology at $\gamma=0$ found in unitary and spatially-flat gauges is
truly a coordinate singularity, one has to prove that in Newtonian gauge not only does the gauge variable $\sigma$ remain finite at $\gamma=0$ but all the other gauge variables remain non-singular as well. 

\subsection{$\gamma$-crossing in Newtonian gauge}

In Newtonian gauge ($\sigma \equiv 0$), the line element with scalar perturbations is given by 
\begin{eqnarray}
ds ^2 & = & - (1 + 2\Phi) dt^2 + a ^2(t)  \left(1- 2 \Psi \right) \delta _{ij} dx ^{i} d x ^{j} 
\,,
\end{eqnarray}
where the Bardeen potentials 
\begin{eqnarray}
\Phi &\equiv& \alpha - \dot{\sigma}\,, \\
\Psi &\equiv& \psi + H\sigma
\end{eqnarray}
are gauge-invariant.
As above, to characterize perturbations of the scalar field, we use the gauge variable 
\begin{eqnarray}
\delta u & \equiv & - \frac{\delta \phi}{\dot{\phi} }
\;.
\end{eqnarray} 
Again, this means we only consider solutions with $\dot{\phi}\neq 0$. We can do this, though, without loss of generality since  $\dot{\phi}\neq 0$ around $\gamma$-crossing.

Evaluating Eqs.~(\ref{0-0-u-simple}-\ref{diag-u-simple0}) and \eqref{scalar-eq-u0} in Newtonian gauge, the linearized Einstein field equations are
\begin{eqnarray} 
\label{0-0-newt}
& & \left( 6 H\gamma(t) - 3A _h(t)H^2(t) - \rho_K(t)  \right) \Phi + 3\gamma(t)\dot{\Psi} +  \frac{k^2}{a^2} A_h(t) \Psi  
\qquad\\
&+&  \left( 3H(t)\gamma(t) -3A_h(t)H^2(t)  - \rho_K(t) \right)  \delta\dot{u} 
 +   \left( -3\dot{H} \gamma(t) + \frac{ k^2}{a^2}\Big(\gamma(t) -A_h(t)H(t) \Big)  \right) \delta u 
= 0
\,,\nonumber\\
\label{t-i-newt} 
& & \gamma(t)\Phi + A_h(t) \dot{\Psi} 
-  \Big( A_h(t)H(t) - \gamma(t)  \Big) \delta \dot{u} 
- A_h(t) \dot{H}  \delta u =0
\,,\\
\label{offdiag-u-newt}
&& A_h(t)  (\Phi - \Psi)  =  \dot{A}_h(t) \delta u 
\,, 
\\
\label{diag-u-simple-newt}
& & \gamma(t)\dot{\Phi} +  \left( \dot{\gamma}(t) + 3H\gamma(t) \right)\Phi +
A_h(t)\ddot{\Psi}  +  \left(3A_h(t)H + \dot{A}_h(t) \right) \dot{\Psi} 
\\
&-&  \left( A_h(t)H(t) - \gamma(t)  \right)  \delta\ddot{u}  
+  \left(  \dot{\gamma}(t) - 3H\left( A_h(t)H(t) - \gamma(t)  \right) - 2A_h(t)\dot{H}(t) - \dot{A}_h(t)H(t)
 \right) \delta\dot{u} 
\nonumber\\
&-&  \left( A_h\ddot{H} + \dot{A}_h\dot{H}   + 3A_hH\dot{H} \right)\delta u
 = 0 \,;
 \nonumber
\end{eqnarray}
and the scalar-field equation is given by 
\begin{eqnarray}
\label{scalar-eq-u}
&&  \left( \rho_K + 3H \big(A_hH - \gamma\big) \right) \dot{\Phi} -\big( A_hH - \gamma \big)\frac{k^2}{a^2}\Phi
\\
&+& \left( \dot{\rho}_K + 3H\rho_K + ( 6\dot{H} + 9H^2) \big(A_hH - \gamma \big)  
+  3H \big( \dot{A}_hH -\dot{\gamma} \big) \right) \Phi
\nonumber\\
& + &  3\big(A_h(t)H(t) - \gamma(t)\big) \ddot{\Psi}   - \dot{A}_h(t) \frac{k^2}{a^2}  \Psi 
+ 3 \left(  3H(t) \big(A_h(t)H(t) - \gamma(t)\big) + \dot{A}(t)H(t) -\dot{\gamma}(t) \right) \dot{\Psi}
\nonumber\\
&+&   \rho_K(t) \delta\ddot{u} + \left( \dot{\rho}_K(t)  + 3H \rho_K(t)  \right) \delta\dot{u} 
+  \Big( H\big(A_h(t)H(t) - \gamma(t)\big) + 2\dot{A}_hH  -\dot{\gamma}  \Big)  \frac{k^2}{a^2}\delta u
\nonumber\\
&-&  3\left( \big(\dot{A}_hH - \dot{\gamma}\big)\dot{H} + \big(A_hH - \gamma\big) \big(\ddot{H} + 3 H\dot{H}\big)\right)  \delta{u}  
= 0\,.
\nonumber
\end{eqnarray}

It is easy to see that Eqs.~(\ref{0-0-newt}-\ref{offdiag-u-newt}) form a closed system for the three scalar gauge variables $\Phi, \Psi$, and $\delta u$ iff $A_h \neq 0$:
Notably, since the Newtonian gauge condition removes the only dynamical variable from the anisotropy equation~\eqref{offdiag-u-newt}, it reduces to an algebraic relation, such that we can use it to eliminate one of the three gauge variables. 
Substituting Eq.~\eqref{offdiag-u-newt}, {\it e.g.}, for the Newtonian potential  $\Phi$ in the Hamiltonian and momentum constraints and putting the resulting linear system into matrix form
\begin{equation}
P(t)\left(\begin{array}{c}\dot{\Psi} \\ \delta \dot{u} \end{array}\right) = Q(t)  \left(\begin{array}{c}\Psi \\ \delta u \end{array}\right) \,,
\end{equation}
where 
\begin{equation}
P(t) = \left(\begin{array}{cc}
3\gamma \quad & 3H\gamma - 3A_hH^2 - \rho_K  \\
A_h \quad & \gamma - A_hH
\end{array}\right) ,
\quad
Q(t) = \left(\begin{array}{cc}
q_{11} & 
q_{12}  \\
-\gamma & q_{22}
\end{array}\right) ,
\end{equation}
with the matrix elements being given by
\begin{eqnarray}
q_{11}(t) &=&  \rho_K + 3 A_hH^2 - 6 H\gamma -\frac{k^2}{a^2} A_h  
\,,\\
q_{12}(t) &=&   3\dot{H}\gamma  + \frac{\dot{A}_h}{A_h}\left( \rho_K + 3 A_hH^2 - 6H\gamma\right) + \frac{k^2}{a^2}(A_hH-\gamma ) \\
q_{22}(t) &=& A_h\dot{H} - \frac{\dot{A}_h}{A_h}\gamma\,,
\end{eqnarray}
it becomes manifest that the system is non-singular at $\gamma$-crossing: the determinant of the kinetic matrix,  
\begin{eqnarray}
\det(P) &=& A_h\rho_K +  3\left(A_hH - \gamma  \right)^2  \equiv r(t)
\end{eqnarray}
is non-zero for all times $t \sim t_{\gamma}$, if $A_h\neq 0$. We note that the positivity condition on $\det(P)$ is equivalent to the no-ghost condition.

The Newtonian gauge results can be verified by using a differential gauge -- harmonic gauge; see the Appendix~\ref{app-harmonicEqs}.  Here too we find non-singular behavior around gamma-crossing.  However, it is inconvenient to use harmonic gauge in general.  Although the harmonic gauge is commonly implemented in numerical general relativity codes due to its stability characteristics, it is more difficult to interpret.  The Newtonian gauge has the advantage that it enables an explicit analytic calculation of the perturbation variables and gives an intuitive understanding of the stability behavior around $\gamma$-crossing. In addition, the harmonic gauge must yield equivalent results concerning stability because the Bardeen potential $\Psi = \psi_h + H\sigma_h$ is gauge-invariant. That means, if we find in Newtonian gauge that the evolution of $\Psi$ is pathological above a certain wavenumber $k$, the harmonic-gauge variables $\psi_h$ and/or $\sigma_h$ must too be ill-behaved above the same wavenumber $k$.

\section{$\gamma$-crossing: linear stability analysis in Newtonian gauge}
\label{sec:newtonian}

The goal of this section is to perform a linear stability analysis about $\gamma$-crossing of the conformally-coupled ${\cal L}_4$ Horndeski theory as specified by Eq.~\eqref{G-i} in Newtonian gauge. 
What we have demonstrated in Sec.~\ref{sec:3+1} is that unitary and spatially-flat gauges lead to the perturbation equation Eq.~\eqref{v-equation} that is valid as long as the corresponding spatial hypersurfaces remain spacelike and non-singular. However, there is a problem with the gauge condition with Horndeski theories near $\gamma$-crossing because the lapse and shift constraints lead to a coordinate singularity.  
Consequently, even though the sound speed in Eq.~\eqref{v-equation} becomes imaginary at gamma-crossing for all $k$,  indicating an instability at all wavelengths, we have to check whether the instability is physical or an artifact of the gauge choice or slicing.  Here we will show that it is indeed an artifact by switching to a gauge (Newtonian) that is well-behaved at $\gamma$-crossing and solving the linearized Einstein equations for the Newtonian scalar gauge variables. 

We will show that, precisely because of the braiding effect, the {\it sign} of the square of the effective sound speed is $k$-dependent and positive for {\it all} wavelengths (of order the Planck scale or above) at $\gamma$-crossing for Horndeski theories.  This means perturbations pass through $\gamma$-crossing undisturbed.
The result is to be contrasted with perfect fluid type models ({\it i.e.}, $P(X, \phi)$  theories) where the square of the sound speed is scale-independent and becomes negative for all wavelengths during any NEC-violating phase, such as a cosmological bounce.

\subsection{Evolution equation around $\gamma$-crossing}

The advantage of using an algebraic gauge is that the linear stability analysis is equivalent to solving a single second-order differential equation for each co-moving wave-number $k$. 
After eliminating two of the three scalar gauge variables - the scalar velocity  potential $\delta u$ using 
\begin{equation}
  (\dot{A}_h/A_h) \delta u = \Phi - \Psi
\,,
\end{equation}
and the Newtonian potential $\Phi$ using
\begin{align}   
 \left(\dot{H} - \frac{\dot{A}_h}{A_h} H - \frac{ k^2}{a^2} \frac{\left( A_hH - \gamma \right)^2}{\det(P)} \right)\left(\Phi - \Psi  \right) = \frac{\dot{A}_h}{A_h}\left(\dot{\Psi} + H \Psi - \frac{k^2}{a^2} A_h  \frac{A_h H - \gamma }{\det(P) }   \Psi \right)
\end{align}
-- the system of linearized Einstein equations~(\ref{0-0-newt}-\ref{offdiag-u-newt}) reduces to a dynamical equation for the Bardeen potential $\Psi$: 
\begin{equation}
\label{psi-final-eq}
\ddot{\Psi} + F(t, k) \dot{\Psi} + \left( m_0^2(t, k) + c_S^2(t, k)\frac{k^2}{a^2} + u_H^2(t, k)\frac{k^4}{a^4} \right) \Psi=0
\,.
\end{equation}

The coefficient of the friction term $\propto \dot{\Psi}$ is given by
\begin{align}
\label{friction-final}
F(t, k) \equiv &  \Bigg( \det(P) \left( 
\left(H + \frac{\dot{A}_h}{A_h}\right) \left(-\dot{H} + \frac{\dot{A}_h}{A_h} H \right) - \frac{d}{dt}\left(-\dot{H} + \frac{\dot{A}_h}{A_h} H \right) \right) 
\\
& 
+ \left(   \frac{d}{dt} \ln \frac{a^3\, A_h\, \det(P)}{ \left( A_hH - \gamma \right)^2 }  \right) \left( A_hH - \gamma \right)^2 \frac{ k^2}{a^2} \Bigg) \frac{1}{d(t,k)}
\,;\nonumber
\end{align}
in the limit of large and small $k$, $F(t, k)$ is a function of $t$ only and so, generally, does not affect stability. 

The coefficient of the term $\propto \Psi$ is given by
\begin{align}
\label{m_0-final}
m_0^2(t, k) \equiv &  \Bigg(   2\dot{H} - H\frac{d}{dt}\ln \left(-\dot{H} + \frac{\dot{A}_h}{A_h} H \right)
\Bigg) \left(-\dot{H} + \frac{\dot{A}_h}{A_h} H\right)  \frac{ \det(P)}{d(t,k)}
\, ,\\
\label{c_S-final}
c_S^2(t, k) \equiv& \Bigg(  \left(-\dot{H} + \frac{\dot{A}_h}{A_h} H \right)\left(  \det(P)c_{\infty}^2(t) + 2A_h\left( \dot{\gamma} + (A_hH - \gamma)H  - \frac{d}{dt}(A_hH) \right) \right) 
\\
&+ 2(\dot{H}+H^2)(A_hH - \gamma)^2 + A_h(A_hH - \gamma) \frac{d}{dt} \left(-\dot{H} + \frac{\dot{A}_h}{A_h} H \right) - H \frac{d}{dt} \left( A_hH - \gamma \right)^2
\nonumber\\
&+  \left(A_h(A_hH - \gamma) (-\dot{H} + \frac{\dot{A}_h}{A_h} H) + H(A_hH - \gamma)^2\right) \frac{d}{dt}\ln \det(P) 
\Bigg)\frac{1}{d(t,k)} 
\,,\nonumber\\
\label{u_H-final}
u_H^2(t, k) \equiv&   \frac{1}{d(t,k)} \left( A_hH - \gamma \right)^2 c_{\infty}^2(t) 
\,.
\end{align}
Notice that all coefficients share a common denominator
\begin{equation}
\label{def-d}
d(t, k) = 
\det(P)\left(-\dot{H} + \frac{\dot{A}_h}{A_h} H \right)  +  \left( A_hH - \gamma \right)^2\frac{ k^2}{a^2}\,.
\end{equation}
Finally, the quantity
\begin{equation}
\label{c_s^2}
c_{\infty}^2(t) \equiv \frac{2\dot{A}_h\gamma  + (A_hH-\gamma)\gamma -A_h\dot{\gamma} }{\det(P)}
\end{equation}
is the square of the propagation speed in the limit of $k\to \infty$. 

Comparing to Eq.~\eqref{v-equation}, it is obvious that $c_{\zeta}^2(t) =  c_{\infty}^2(t)$. 
Of course, as we have shown above, the gauge variable $\zeta$ is ill-defined around $\gamma$-crossing. That is the real cause of its blow-up, not the fact that $c_{\zeta}^2(t)<0$, as was believed previously. Consequently, we cannot directly compare the behavior of $\zeta$ (or $v$) and $\Psi$ around $\gamma$-crossing. But we can still ask if the imaginary sound speed associated with $\gamma$-crossing in unitary gauge indicates a real physical problem such that it spoils the Newtonian gauge evolution. What we will show is that, even though $\Psi$ has the same sound speed in the limit $k \to \infty$, as noted above, all {\it macroscopic modes} propagate at a different speed $c_S$ that is real. In the remainder of this paper, we use the term `macroscopic' to refer to modes with wavelength greater than the Planck length.
 That means, what looked to be a sickness in unitary gauge is not in Newtonian. 
As a corollary we recover that, in general, $\zeta$ as defined in unitary gauge to be a  local perturbation to the scale factor $a(t)$ does {\it not} characterize the behavior of co-moving curvature modes \cite{Kobayashi:2010cm}. Hence, the criterion that $ c_{\infty}^2(t)>0$ is, in general, not the proper condition for linear stability.\footnote{
It has been a common claim in the literature that stability requires $c_{\infty}^2(t)>0$; see, {\it e.g.}, \cite{Bellini:2014fua}. In reality, this condition is necessary only if one demands strong hyperbolicity of the classical linearized equation not just for macroscopic modes but also for wavelengths smaller than the Planck scale; see, {\it e.g.}, \cite{Babichev:2007dw}. For all practical purposes, such a condition is too strong.
}

Because the friction term $F(t,k)$ is weakly $k$-dependent and has no effect on stability, as we argued above, the effective sound speed is determined by the coefficient of the $\Psi$ term.  The fact that this coefficient includes terms $\propto k^2$ and $\propto k^4$ implies that the sound speed is generally $k$-dependent.  This renders the question of linear stability in Horndeski theories more complex than in the case of perfect-fluid type forms of stress-energy. 
Most importantly, the non-trivial $k$-dependence in the $\Psi$ equation~\eqref{psi-final-eq} introduces two additional scales entering the dynamics: 
The first scale corresponds to the wavenumber $k_B$ that divides the range of $k\lesssim k_B$ where the homogeneous term dominates in the denominator $d(t,k)$ from the range $k_B \lesssim k$ where the $k$-dependent term dominates:
\begin{equation}
\label{k_T}  
\left( \frac{ k_B}{a} \right)^2= \frac{\det(P)}{ \left( A_hH - \gamma \right)^2}\left|-\dot{H} + \frac{\dot{A}_h}{A_h} H \right|  \,.
\end{equation} 
We note that this novel type of scale-dependence was previously identified in the dark-energy literature where $k_B/a$ was dubbed the `braiding scale' \cite{Bellini:2014fua}. Interestingly, none of these two scales have been noticed in the context of stability consideration of early-universe scenarios. 
We suspect they were overlooked because in the commonly used unitary and spatially-flat gauges the braiding between the metric and the scalar field is not apparent due to the fact that either the metric nor the scalar degrees of freedom are set to be non-dynamical.

The second scale, which is more important is given by the ratio
\begin{equation}
\label{k_B}  
\left( \frac{ k_T}{a} \right)^2 = \frac{c_S^2}{u_H^2}  \,;
\end{equation} 
we call this scale the `transition scale' because it defines the scale where the dynamics transitions from being dominated by the $k^2$-term to being dominated by the $k^4$-term in Eq.~\eqref{psi-final-eq}. In general, the two scales do not coincide. In fact, we will see that around $\gamma$-crossing the transition scale can lie well above the braiding scale.

To characterize the generic stability behavior around $\gamma$-crossing (where $|t-t_{\gamma}|$ is small), we approximate the quantities $a(t), \gamma(t)$ and $A_h(t)$ using their Taylor expansion:
\begin{eqnarray}
\label{taylor-gamma}
\gamma(t) &=& \gamma_0 \Delta t
\,,\\
a(\Delta t) &=& \left(1+\frac{\Delta t }{t_{\gamma}}\right)^{p} 
\,,\\
\label{taylor-Ah}
A_h(\Delta t) &=& A_0 + A_1 (\Delta t)^2
\,.
\end{eqnarray}
Here 
\begin{equation}
\Delta t \equiv t-t_{\gamma}
\end{equation}
measures the time that passed since $\gamma$-crossing $t_{\gamma}$; and all parameters $A_0, A_1, p$, and $\gamma_0$ are positive real constants. For clarity, we assume a slowly contracting (ekpyrotic) phase ($p\lesssim 1/3$) well before the NEC-violating bounce stage (including $\gamma$-crossing at $t_{\gamma}<0$), {\it i.e.}, $t<0$ runs towards zero. 
It has been shown in Ref.~\cite{Ijjas:2016vtq} that, for generic $\gamma$-crossing, $A_h$ has to be a function of time near $\gamma=0$, corresponding to the conformal ${\cal L}_4$ coupling. Here, for simplicity, we chose $A_h$ to have its minimum at $\gamma=0$. 
In addition, to keep all physical scales well below the Planck scale, we require 
$|t_{\gamma}|  \gg  1; |H|, |\gamma | \lesssim 10^{-3}$; and we normalized $a(t)$ so that $a(t_{\gamma})=1$.
Our choice of coefficients ensures that $A_h(\Delta t)$ is strictly positive throughout. To avoid strong-coupling issues related to letting Newton's constant approach infinity, here we will only consider values $0 \ll A_h(\Delta t) \lesssim 1$. 

To express the kinetic-matrix determinant 
\begin{eqnarray}
\det(P) = A_h\Big(  \frac12 \kappa(\phi)\dot{\phi}^2   + \frac12 \big(3q(\phi) - 2b_{,\phi}  \big) \dot{\phi}^4 + 3Hb(\phi)\dot{\phi}^3\Big) + 3\Big(A_hH-\gamma\Big)^2
\end{eqnarray}
as a function of $A_h, H$, and $\gamma$, we have to fix the functional form of $G_2(X, \phi)$. We take the most conservative approach and  choose a coupling with minimum number of degrees of freedom, {\it i.e.},
\begin{equation}
G_2(X,\phi)=\kappa(\phi)X+q(\phi)X^2 - V(\phi);
\end{equation}
such that, using the background equations~(\ref{FRW1}-\ref{FRW2}), we can eliminate  both $\kappa(\phi)$ and $q(\phi)$,
\begin{eqnarray}
\label{kappa-def}
 \kappa(\phi)\dot{\phi}^2 &=& -2\Big(\dot{\gamma} + 3H\gamma + 2\dot{A}_hH + A_h\big( 2\dot{H} + 3H^2\big) + \ddot{A}_h- 2V \Big)
\,,\\
\label{q-def}
q(\phi)  \dot{\phi}^4 &=& \frac43 \Big(\dot{\gamma} + 9H\gamma + 2\dot{A}_hH + 2A_h\dot{H} + \ddot{A}_h- 3V \Big) + \frac23 b_{,\phi} \dot{\phi}^4\,.
\end{eqnarray}
Obviously, more general forms for $G_2(X, \phi)$ would leave us with residual degrees of freedom and could further ease the stability constraints.
Substituting Eqs~(\ref{kappa-def}-\ref{q-def}) into the expression for $\det(P)$ yields
\begin{equation}
\det(P) = A_h\Big(  \dot{\gamma} + 9H\gamma + 5\dot{A}_hH + 2A_h\dot{H} + 3A_hH^2+\ddot{A}_h-4V\Big) + 3\Big(A_hH-\gamma\Big)^2
\,.
\end{equation}

\subsection{Analytic approximation}
\label{subsec:analytic}

Even though the coefficients of Eq.~\eqref{psi-final-eq} may appear to be complicated, it is actually straightforward to analytically approximate them and gain an intuitive understanding of the dynamics in the vicinity of $\gamma$-crossing, for details of the derivation see Appendix~\ref{sec:approx}.
First, we note that the kinetic-matrix determinant is nearly constant 
\begin{equation}
\det(P) \simeq (-4A_0 V_0)\,,
\end{equation}
and the expression for the `braiding scale' is well-approximated by 
\begin{equation}
\label{def-k_B}
\frac{k_B(\Delta t)}{a} \simeq  \sqrt{\frac{(-4 V_0)}{A_0 p}} \left( 1 + \gamma_0 \frac{t_{\gamma}^2 }{A_0 p}  \left(- \frac{\Delta t}{t_{\gamma}}\right) \right)^{-1}
\,.
\end{equation}
Remarkably, even with $A_0 \sim 1$, the braiding scale $k_B \sim 1$ if $(-4V_0/p) \sim 1$. Parameters in this range are typical, for example, for a slowly contracting phase before entering the bounce stage when the scalar moves uphill a negative potential. 
This means, restricting our analysis to modes with $k\lesssim k_B\sim 1$ spans all modes with wavelengths larger than the Planck length, {\it i.e.}, all modes that can be reliably tracked using classical equations of motion.
As we will see, around $\gamma$-crossing the transition scale typically lies well above the braiding scale. 
Hence, over the range of wavelengths $k\lesssim k_B\sim 1$, the relevant quantity that determines stability is $c_S^2$ which differs significantly from $c_{\infty}^2$; consequently, the sign of $c_{\infty}^2$ is {\it not} decisive in determining stability in early-universe scenarios involving NEC violation.\footnote{
Note that this situation is substantially different from applications of Horndeski theories to explain dark energy in the late universe. In this case  $k_{\rm B}$ is order of the horizon scale rather than the Planck scale. The modes of interests are on subhorizon wavelengths $k\gg k_B$ where the effective sound speed is $c_{\infty}$ so the sign of  $c_{\infty}^2$ is decisive in determining stability.
}

In the limit $k\lesssim k_B \simeq 1$, the common denominator of the expressions $\propto \dot{\Psi}, \Psi$ is approximately $k$-independent, and near $\gamma$-crossing roughly constant:
\begin{eqnarray}
d(\Delta t, k) 
& \simeq& \frac{p }{ t_{\gamma}^2}(-4A_0V_0) \left( 1  + 2  \left( 1 - \frac{A_1}{A_0}
  t_{\gamma}^2\right) \left(-\frac{\Delta t}{t_{\gamma} }\right) +  \left(3 - 2 \frac{A_1}{A_0} t_{\gamma}^2  \right)\left(-\frac{\Delta t}{t_{\gamma} }\right)^2  \right) 
 \quad \\ 
 & \simeq&  \frac{p }{ t_{\gamma}^2}(-4A_0V_0) \nonumber
  \,,
\end{eqnarray}
such that the $\Psi$ equation~\eqref{psi-final-eq} takes the simplified form
\begin{equation}
\label{psi-approx}
\ddot{\Psi} + F(\Delta t) \dot{\Psi} + \Big( m_0^2(\Delta t) + c_S^2(\Delta t)k^2 - u_H^2(\Delta t)k^4 \Big) \Psi=0\,.
\end{equation}
Now, the coefficient term of ${\Psi}$ is given by the time-dependent expressions
 \begin{eqnarray}
 \label{m_0-ap}
 m^2_0 (\Delta t) & \simeq & -2 p\frac{A_1}{A_0 }
\,,\\
\label{c_S-ap}
c^2_S(\Delta t) &\simeq& \frac{1}{(-4V_0)}  \left(1 - 2 \frac{A_1}{A_0}t_{\gamma}^2 \left( - \frac{\Delta t}{ t_{\gamma} } \right)  \right) \Big((-4V_0)c_{\infty}^2(t)  + 2\gamma_0 \Big) 
 \\
&+&  \frac{2}{(-4V_0) } \left( p  +  \frac{\gamma_0 t_{\gamma}^2}{A_0} \frac{\Delta t}{(-t_{\gamma})} \right) \gamma_0
\nonumber\\
&+&  \frac{2A_0}{(-4V_0) (-t_{\gamma})^2}  \left( \frac{A_1}{A_0}t_{\gamma}^2 \right) \left(1 - 2  \left( - \frac{\Delta t}{ t_{\gamma}} \right)  \right) \left( p +\frac{\gamma_0 t_{\gamma}^2}{A_0}  \frac{\Delta t}{(- t_{\gamma})}  \right)
\nonumber\\
 &+& \frac{2A_0}{(-4V_0)} \frac{p}{(-t_{\gamma})^2} \left(  1+p + \frac{\gamma_0 t_{\gamma}^2 }{A_0}\frac{\Delta t}{(-t_{\gamma})}\right) \left(1 - 2 \frac{A_1}{A_0}t_{\gamma}^2 \left( - \frac{\Delta t}{ t_{\gamma}} \right)  \right)  
\nonumber\\
&-& \frac{2 A_0}{(-4V_0) }  \frac{(1-p)}{(-t_{\gamma})^2} \left( p  +  \frac{\gamma_0 t_{\gamma}^2}{A_0} \frac{\Delta t}{(-t_{\gamma})} \right)^2
\,,  \nonumber\\
\label{u_H-ap}
u_H^2 (\Delta t) & \simeq&  \frac{A_0 }{(-4V_0)p } \left( p +  \frac{\gamma_0 t_{\gamma}^2}{A_0}\frac{\Delta t}{(-t_{\gamma})}  \right)^2 
c_{\infty}^2(t) 
 \,.
\end{eqnarray}
\begin{figure}
\centering
\includegraphics[width=8.5cm]{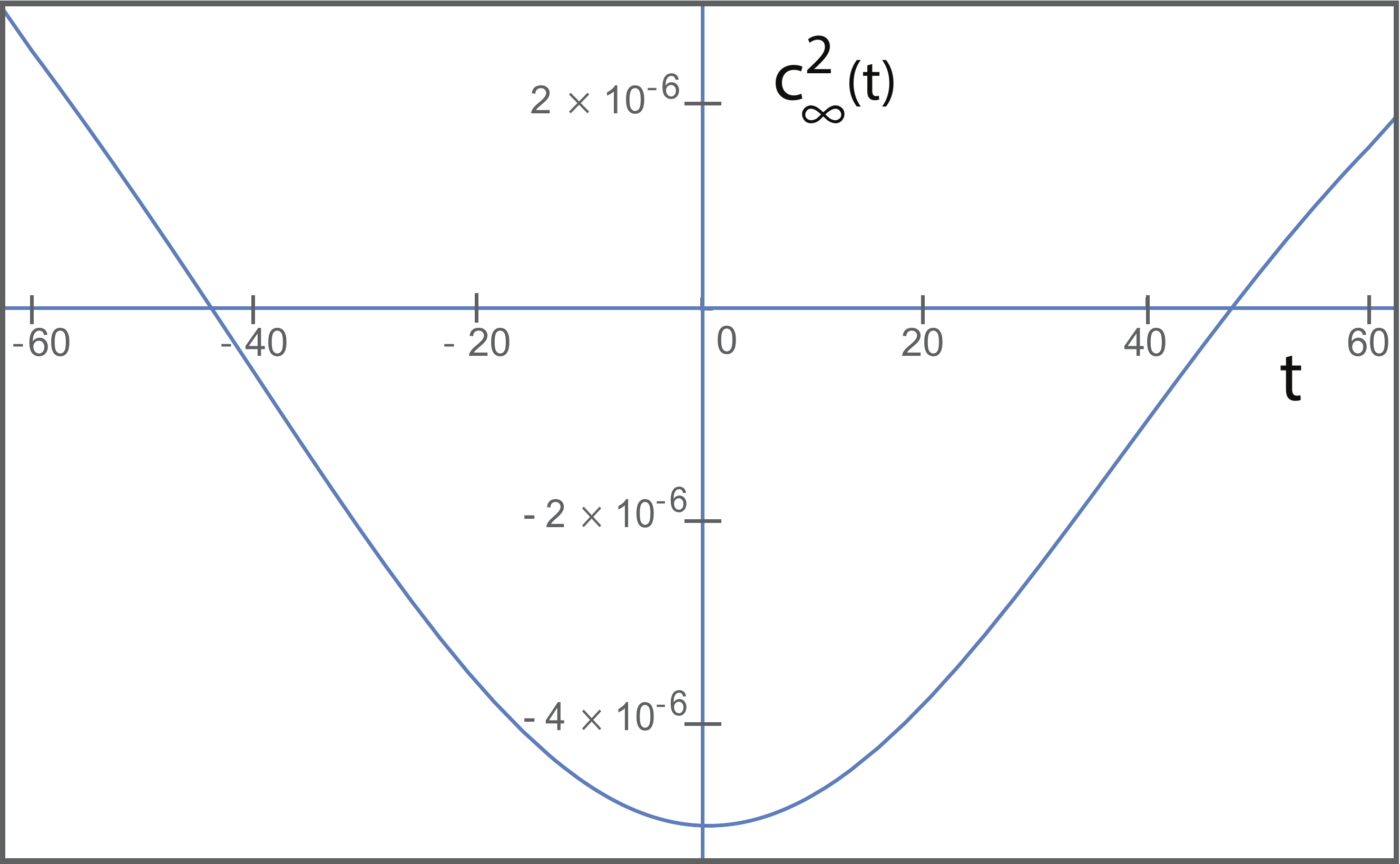}
\caption{Evolution of $c_{\infty}^2$ around $\gamma$-crossing as a function of time for the parameter values given in the caption of Figure~2.
The time coordinate is given in reduced Planck units; the $y$-axis has dimensionless units. The $\gamma$-crossing point is at $t=0$. Notice that, even though $|c_{\infty}^2| \ll 1$ around $\gamma$-crossing in this example, the true effective sound speed $c_S^2$ is generally much greater than  $|c_{\infty}^2|$ and can be of order one.}
\label{fig1}
\end{figure}
where $c_{\infty}^2(\Delta t)$ can be approximated as
\begin{equation}
c_{\infty}^2(\Delta t) \simeq \frac{\gamma_0}{(-4 V_0)} \left( -1 + 3 \frac{A_1}{A_0} (-t_{\gamma})^2 \left( \frac{\Delta t}{t_{\gamma}}\right)^2 \right) \simeq - \frac{\gamma_0}{(-4 V_0)} \,.
\end{equation}
 We note that, keeping only terms up to second order in $\Delta t/t_{\gamma}$, differential equations of the form~\eqref{psi-approx} are analytically solvable; the solutions are `parabolic cylinder' or `Weber-Hermite functions.'  More importantly, though, we can read of the dynamical behavior of the modes from the approximate coefficient expressions:

The  term in Eq.~\eqref{psi-approx} relevant for the stability analysis near $\gamma$-crossing is the one proportional to $\Psi$. The friction term $F$ is nearly $k$-independent around $\gamma$-crossing, and so has no bearing on stability, as shown above.  Also note that  $c_S^2>0$ is greater than $|u_H^2|$ as can be seen from Eqs.~(\ref{c_S-ap}-\ref{u_H-ap}).
Given that, the coefficient of $\Psi$ suggests three different regimes of behavior:  (i.) $k \ll k_J \equiv m_0/c_S$; (ii.) $ k_J \ll k \ll k_T \equiv c_S/u_H$; and (iii.) $k_T \ll k$ each of which is associated with a different dynamics:
\begin{figure}%
    \centering
    \includegraphics[width=9.5cm]{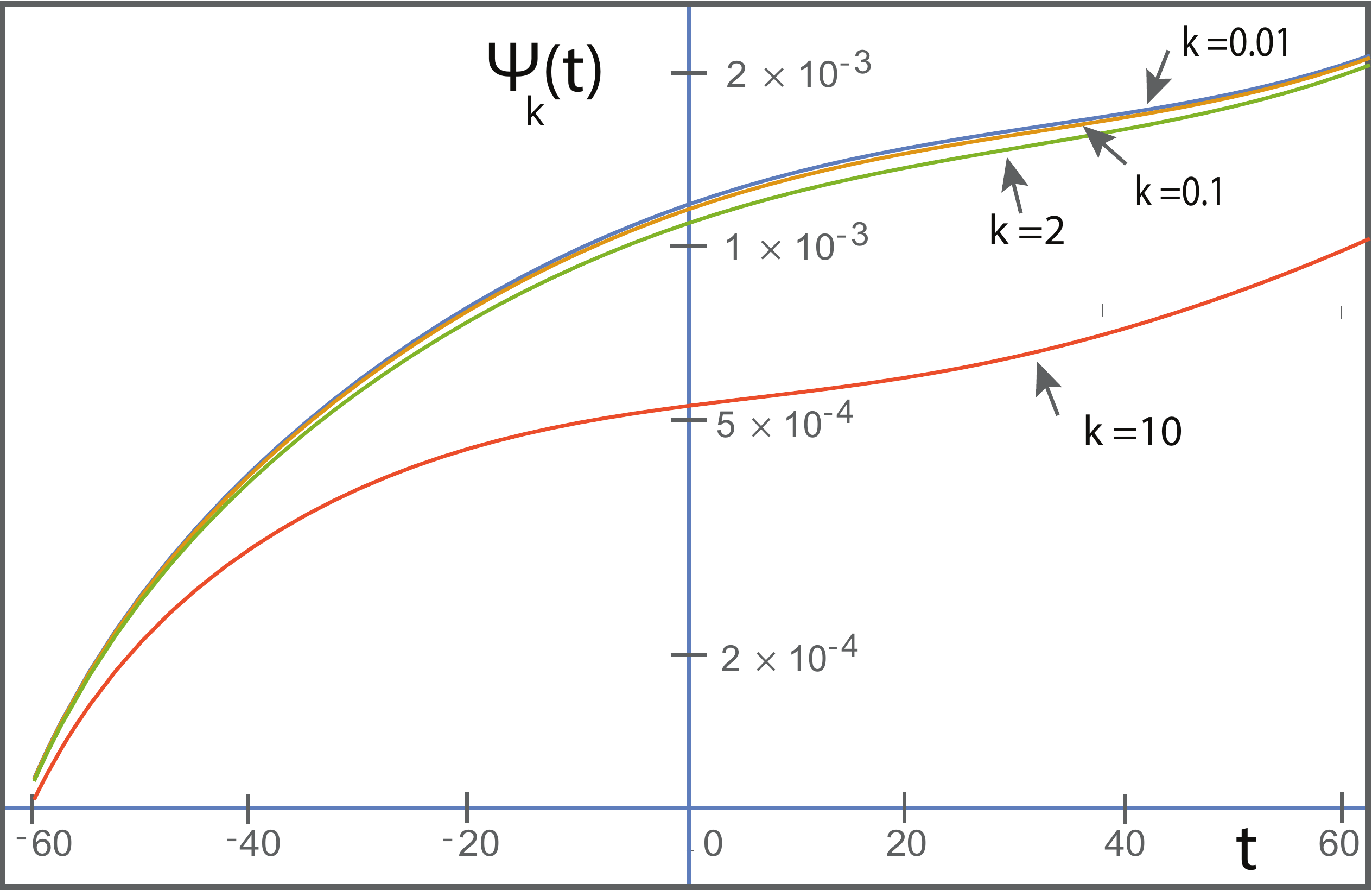} %
    \caption{Numerical solution of Eq.~\eqref{psi-final-eq} for the modes $\Psi_k$ as a function of time corresponding to the background around $\gamma$-crossing as described in Eqs.~(\ref{taylor-Ah}-\ref{taylor-gamma}) for the wave-numbers $k=0.01$ (blue curve), $k=0.1$ (orange curve), $k=2$ (green curve), and $k=10$ (red curve); and the parameters $t_{\gamma}=-100, p=1/10, A_0=10^{-3}, A_1= 2/10, V_0= -5\times10^{-3}$, and $\gamma_0 =10^{-6}$. The $x$-axis has reduced Planck units and the $y$-axis has dimensionless units. The graph verifies that all modes pass through $\gamma$-crossing undisturbed.}%
    \label{fig:two}%
\end{figure}

\begin{enumerate}
\item[i.] if $k \ll k_J$, the evolution of the modes is determined by the homogeneous mass term $m_0^2(t)$. It defines the so-called `Jeans scale.' Since $m_0^2(t)<0$, these ultra-long wavelength modes experience some growth around $\gamma$-crossing. But no significant growth can develop since the typical timescale for it to arise is given by $1/|m_0| \gtrsim {\cal O}(1000)$ (Planck times) that is much longer than the time-scale of $\gamma$-crossing which is $\sim 2|\Delta t| \sim {\cal O}(10)$.
\item[ii.] if $ k_J \ll k \ll k_T $, the modes propagate at the speed $c_S$. It is clear from Eq.~\eqref{c_S-ap} that $c_S^2>0$ for all times $\Delta t$: Intriguingly, the $2\gamma_0$ contribution in the first line of Eq.~\eqref{c_S-ap} compensates for the negative term $\propto c^2_{\infty}$ ($-4V_0c_{\infty}^2(t)  + 2\gamma_0 \simeq \gamma_0>0$) and, manifestly,  the remaining terms together also give a positive contribution.
That means, all modes in this regime exhibit an oscillatory behavior. No instability arises.
\item[iii.] if $k_T \ll k$, the evolution of the modes is determined by $c_{\infty}^2$. Since it is negative around $\gamma$-crossing, modes will grow exponentially according to the classical equations. However, all these modes have wavelengths smaller than the Planck scale. For example,
near $\gamma$-crossing ($\Delta t \simeq 0$), the transition scale is approximately given by
\begin{equation}
\frac{k_T}{a} = \frac{c_S}{u_H} \simeq 2 \sqrt{\frac{(-V_0)}{A_0} \left( \frac{2A_1}{\gamma_0}+ \frac{p+2}{p} \right)}\,.
\end{equation}
Note that the transition scale $k_T > k_B \simeq 2 \sqrt{(-V_0/(A_0 p)} $.
\end{enumerate}

To verify our analytic estimates, we numerically solve Eq.~\eqref{psi-final-eq} for $\Psi$, assuming initial conditions that are common after a long smoothing phase such as ekpyrosis, {\it i.e.}, $\Psi_0 = 10^{-5}, \dot{\Psi}_0 = 10^{-5}$. The results are given in Figure~2.
Notably, even though the Taylor approximation is designed only to describe the behavior for $\Delta t/ t_{\gamma}\ll 1$, the numerical analysis proves that all $k\lesssim {\cal O}(10)$ are well-behaved throughout the time when $c_{\infty}^2<0$. Outside of this region, {\it i.e.} when $c_{\infty}^2>0$, it is straightforward to avoid any instability, as has been shown  in Refs.~\cite{Ijjas:2016tpn,Ijjas:2016vtq}.
From Figure~3, one can see that the analytic approximation is in excellent agreement with the numerical result for the time range $|\Delta t| \lesssim 20$.
\begin{figure}%
    \centering
    \includegraphics[width=9cm]{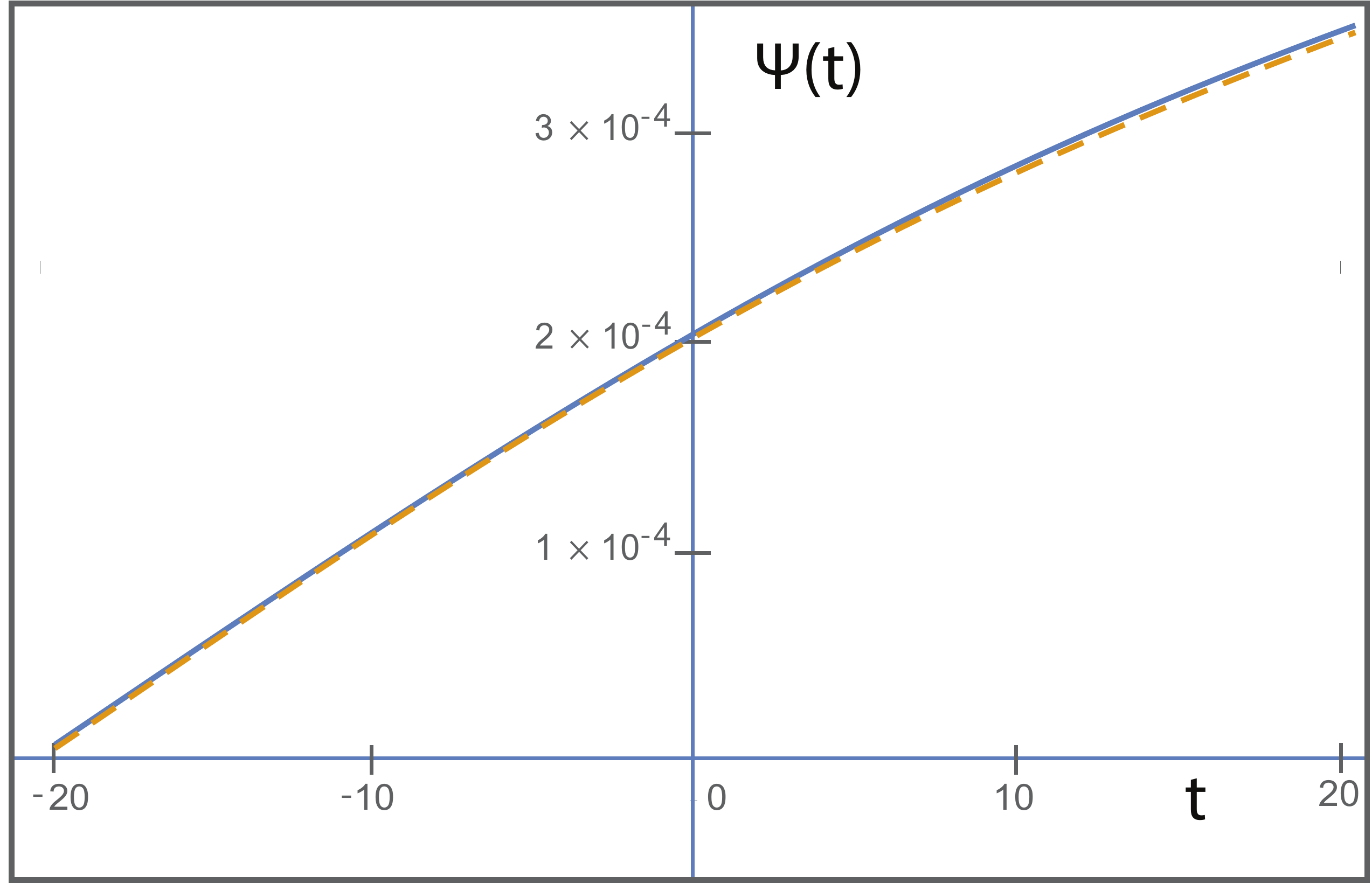} %
    \caption{Comparison of solutions for $\Psi(t, k=1)$ for the exact coefficients (continuous blue curve) as given in Eqs.~(\ref{m_0-final}-\ref{u_H-final}) and the analytic approximation (dashed orange curve) as given in Eqs.~(\ref{m_0-ap}-\ref{u_H-ap}). The $x$-axis has reduced Planck units and the $y$-axis has dimensionless units. The parameters are the same as given in the caption of Figure~2. The graph demonstartes that for the range $|\Delta t| \sim 20$ the exact solution and the analytic approximation are in excellent agreement even at $k=1$.}%
    \label{fig:two}%
\end{figure}

\section{Discussion}
\label{sec:summary}

In this paper, we have argued that it is possible to construct non-singular cosmological solutions that admit a NEC-violating bounce (or genesis) stage and reduce to Einstein gravity both before and after the NEC-violation without producing any pathologies. The key was to show that the evolution all modes with macroscopic wavelengths, {\it i.e.}, wavelengths above the Planck length, is determined by an effective sound speed $c_S$ that stays real throughout.  

 Our finding is a result of a combination of different steps: 
 \begin{itemize} 
 \item[-] as our starting point, we heavily relied on Refs.~\cite{Ijjas:2016tpn,Ijjas:2016vtq} that identified the $\gamma$-crossing point as the only source of complications, where  $\gamma$-crossing can occur arbitrarily long before or after the NEC-violating bounce (or genesis) stage;
 \item[-] we demonstrated that the apparent divergence around $\gamma$-crossing found in earlier calculations is actually a coordinate singularity of the commonly used unitary and spatially-flat gauges that were also utilized in deriving all the no-go theorems; 
 \item[-] we showed that the Newtonian gauge is the unique algebraic gauge that avoids the coordinate singularity around $\gamma$-crossing;
 \item[-] performing the stability analysis in Newtonian gauge, we found that braiding between the metric and the scalar field leads to a dynamical equation for the Bardeen potentials with $k$-dependent sound speed. 
 The fact that the $k^2$-term has the correct sign for stability is especially noteworthy. In particular, it is a significant improvement to the commonly considered NEC-violating $P(X,\phi)$ theories, also called `ghost condensate,' where the term $\propto k^2$ immediately becomes imaginary upon entering NEC violation and remains so throughout the NEC-violating phase ($c_S^2 \sim -\dot{H}<0$), preventing macroscopic modes on subhorizon scales from passing through safely. 
 \end{itemize}

Our work opens up several avenues for future work. To mention only a few: 
\begin{itemize}
\item[-] it will be interesting to see how our results affect stability in NEC-violating {\it multifield} bouncing (or genesis) scenarios and whether it has implications for wormhole solutions. In particular, it is necessary to revisit the no-go theorems in Refs.~\cite{Kolevatov:2016ppi,Akama:2017jsa} that utilized the unitary and spatially-flat gauges;
\item[-] similarly,  it will be important to perform a careful computation of the strong coupling and cut-off scales around $\gamma$-crossing.  Commonly, such analyses are based on computing the third-order action in unitary or spatially-flat gauges but, as we have shown, those gauges suffer from a coordinate singularity around $\gamma=0$.
Here, we note that our solutions avoid the usual problem encountered in strong-coupling analyses: namely we find that $c_S$ need not be particularly small at any point during the evolution including $\gamma$-crossing. For example, using Eq.~\eqref{c_S-ap}, we can  approximate the value of the effective sound speed around $\gamma = 0$,
\begin{equation}
c_S \simeq \sqrt{\frac{A_1  p}{(-2V_0)}  }\,,
\end{equation}
where for typical values of $V_0 \sim {\cal O}(10^{-3}), A_1 \sim {\cal O}(10^{-3}) $ and $p \lesssim 1/3$, $ c_S\sim {\cal O}(10^{-1})$. 
We also note that, since $\gamma$-crossing  can occur arbitrary long before NEC violation, {\it i.e.}, at any energy density scale $\sim H^2$, it can be chosen to accommodate any cut-off constraints;
\item[-] finally, since $\gamma$-crossing is a generic feature of all non-singular cosmological solutions that reduce to Einstein gravity both before and after the NEC-violating bounce (or genesis) stage, it will be important to see whether $\gamma$-crossing has distinguishing {\it observational consequences} that could confirm or eliminate these theories empirically.
\end{itemize}

\section*{Acknowledgements}

I thank Claudia de Rham, Rehan Deen, Luis Lehner, Jean-Luc Lehners, Burt Ovrut, Vasileios Paschalidis, Harvey Reall, Valery Rubakov, Andrew Tolley, and Neil Turok for discussions and comments about issues related to this work. I would also like to thank Frans Pretorius for many useful conversations on issues related to gauge choices. I am especially grateful to Paul Steinhardt for all the countless discussions and his very helpful comments on the manuscript.
This research was supported by the Simons Foundation Program `Bouncing Cosmologies and Cosmological Bounces,'  as part of the `Origins of the Universe' Initiative.

\newpage 
\appendix

\section{Linearized conformally-coupled ${\cal L}_4$ Horndeski without gauge fixing}
\label{app-cov-Eqs}

Here, we derive the non-gauge fixed, linearized Einstein equations~(\ref{0-0-u-simple}-\ref{diag-u-simple0}) corresponding to the conformally-coupled ${\cal L}_4$ Horndeski action as given in Eq.~\eqref{G-i} for the scalar gauge variables $\alpha, \beta, \psi, \epsilon$, and $\pi$ specified in Eq.~\eqref{adm-lin-met}.

In terms of these gauge variables, the linearized Einstein tensor $\delta G^{\mu}_{\nu}$ is \begin{eqnarray}
\delta G^0_0 & = & 2 \left( 3 H\left( \dot{\psi} + H \alpha \right) - \frac{\nabla^2}{a^2}( \psi +H\sigma) \right) \,, 
\\
\delta G^0 _i & = & 2 \left( - \dot{\psi} - H \alpha \right)_{,i} 
\,,\\
\delta G^i _j & = & 2\left(  H \dot{\alpha} +  \left(2 \dot{H} + 3 H^2 \right) \alpha +  \frac12 \frac{\nabla^2}{a^2} \left( \alpha -\psi - \dot{\sigma} - H\sigma \right) + \ddot{\psi} + 3H\dot{\psi}   \right)\delta^i_j 
\nonumber\\
&-&\Big( \alpha - \psi - \dot{\sigma} - H\sigma \Big)_{,kj} \delta^{ik} \,,
 \end{eqnarray}
and the linearized stress--energy tensor $\delta T^{\mu}_{\nu}$ is 
\begin{eqnarray}
\delta T^0_0  & = & 
2\left( \frac12 G_{2,X} \dot{\phi}^2 + \left(\frac12 G_{2,X X} - b_{,\phi}  \right) \dot{\phi}^4 + 6 b( \phi) H\dot{\phi}^3 - 3H^2G_4(\phi)  -3HG_{4,\phi}\dot{\phi}  \right) \alpha
\qquad 
\\
&-&2\left( G_4(\phi)H - \frac12 b( \phi) \dot{ \phi}^3 + \frac12 G_{4,\phi}\dot{\phi} \right) \left(3\dot{\psi} - \frac{\nabla^2\sigma}{a^2} \right)
 + 2G_4(\phi) \frac{\nabla^2 \psi}{a^2} 
\nonumber\\
&-&   \left( G_{2,X} \dot{\phi} + \big(G_{2,X X} - 2b_{,\phi}  \big) \dot{\phi}^3 + 9 b( \phi) H \dot{\phi}^2 - 3 G_{4,\phi}H \right) \dot{\pi} 
- \left(G_{4,\phi} - b( \phi)\dot{\phi}^2 \right)  \frac{\nabla^2\pi}{a^2} 
\nonumber\\
&+& \left( G_{2, \phi} - G_{2, X \phi} \dot{\phi}^2  - 3 b_{,\phi} H \dot{\phi}^3 + \frac{1}{2} b_{,\phi\phi} \dot{\phi}^4 + 3 H G_{4,\phi\phi}\dot{\phi} + 3 H^2 G_{4,\phi}  \right)\pi 
\,,
\nonumber \\
\delta T^0 _i & = & 2\left( \left(  G_4(\phi) H +  \frac12 G_{4,\phi} \dot{\phi} -  \frac12 b( \phi)\dot{\phi}^3 \right) \alpha + G_4(\phi) \dot{\psi} 
- \frac12 \left(G_{4,\phi} - b( \phi) \dot{\phi}^2 \right) \dot{\pi} \right)_{,i} 
 \\
&-&  \left( \left(G_{2,X} +G_{4,\phi\phi}\right) \dot{\phi} - G_{4,\phi} H + 3b( \phi) H\dot{\phi}^2  - b_{,\phi} \dot{\phi}^3 \right) \pi_{,i} 
 \,,\nonumber
 \\
\delta T^i _j & = & 2\delta^i_j \Bigg\{ 
- \left( G_4(\phi) H + \frac12 G_{4,\phi}\dot{\phi} - \frac12 b( \phi) \dot{\phi}^3 \right) \dot{\alpha} 
\\
&&  \;
-  \left( G_4(\phi) \left(2 \dot{H} + 3 H^2 \right) + \Big( \frac12 G_{2,X} + G_{4,\phi\phi}\Big) \dot{\phi}^2  -  b_{,\phi} \dot{\phi}^4 + G_{4,\phi}\left(\ddot{\phi} + 2H\dot{\phi} \right)
- 2b(\phi)  \ddot{\phi} \dot{\phi}^2
\right)\alpha
\nonumber\\
&& \;
- G_4(\phi)(\ddot{\psi} + 3H\dot{\psi}) - G_{4,\phi}\dot{\phi} \dot{\psi} + \frac12 \left(G_{4,\phi} - b( \phi) \dot{\phi}^2\right) (\ddot{\pi} + 3H\dot{\pi}) \nonumber\\
&& \; 
+ \frac12 \left( \left( G_{2,X} + 2G_{4,\phi\phi} \right) \dot{\phi} - G_{4,\phi}H - 2 b_{,\phi}\dot{\phi}^3 - b( \phi) \left( 2 \ddot{\phi} -3 H \dot{\phi} \right) \dot{\phi} \right) \dot{\pi}
\nonumber\\
&&  \; 
+ \frac12 \left(G_{2,\phi} - b_{,\phi} \dot{\phi}^2 \ddot{\phi} - \frac{1}{2} b_{,\phi\phi} \dot{\phi}^4 + G_{4,\phi\phi\phi}\dot{\phi}^2 + G_{4,\phi\phi}\left(\ddot{\phi} + 2 H\dot{\phi}\right) 
+G_{4,\phi}(2\dot{H} + 3H^2) \right)\pi
\nonumber\\
&&  \;
 - \frac12 \frac{\nabla^2}{a^2}\Big(  G_4(\phi)\left(  \alpha - \psi  -\dot{\sigma} - H\sigma \right) + G_{4,\phi}\left( \pi - \dot{\phi}\sigma\right) \Big) 
\Bigg\} 
\nonumber\\
&+& \delta ^{ik}  \Bigg\{ G_4(\phi)\left(  \alpha - \psi  -\dot{\sigma} - H\sigma \right) + G_{4,\phi}\left( \pi - \dot{\phi}\sigma \right)  \Bigg\}_{,kj} 
\,.\nonumber
\end{eqnarray}

Combining these expressions, the linearized Einstein equations for a single Fourier mode with wavenumber $k$ take the following form:
\begin{eqnarray}
\label{0-0}
& & 
\left(3 \big(1 + G_4(\phi)\big) H^2 -\frac{1}{2}G_{2,X} \dot{\phi}^2 -\frac{1}{2} \big(G_{2,X X} - 2b_{,\phi}  \big) \dot{\phi}^4 -6 b( \phi) H \dot{\phi}^3  + 3 HG_{4,\phi}\dot{\phi}  \right) \alpha
\qquad\\
&+&  \left( \big(1 + G_4(\phi)\big)H - \frac12 b( \phi) \dot{\phi}^3 + \frac12 G_{4,\phi}\dot{\phi}   \right) \left(3\dot{\psi} +\frac{k^2}{a^2}\sigma \right)
+ \big(1+ G_4(\phi)\big) \frac{k^2}{a^2}  \psi
\nonumber\\
&+&   \frac{1}{2} \left( G_{2,X} \dot{\phi} + \big(G_{2,X X} - 2b_{,\phi}  \big)  \dot{\phi}^3 + 9 b( \phi) H\dot{\phi}^2 - 3 G_{4,\phi}H \right) \dot{\pi} 
- \frac{1}{2}\left(G_{4,\phi} - b( \phi)\dot{\phi}^2 \right) \frac{ k^2}{a^2} \pi 
\nonumber\\
&-&  \frac{1}{2} \left( G_{2, \phi} - G_{2, X \phi} \dot{\phi}^2 - 3 b_{,\phi} H \dot{\phi}^3  + \frac{1}{2} b_{,\phi\phi} \dot{\phi}^4 + 3 H G_{4,\phi\phi}\dot{\phi} + 3 H^2 G_{4,\phi}  \right)\pi 
= 0
\,,\nonumber\\
\label{t-i} 
& &  \left(  \big(1 + G_4(\phi)\big)H+ \frac12 G_{4,\phi} \dot{\phi} - \frac12 b( \phi)\dot{\phi}^3 \right) \alpha + \big(1 + G_4(\phi)\big) \dot{\psi} \\
&-& \frac12 \left(G_{4,\phi} - b( \phi) \dot{\phi}^2 \right) \dot{\pi} 
- \frac{1}{2} \left( \left(G_{2,X} +G_{4,\phi\phi}\right) \dot{\phi} - G_{4,\phi} H + 3b( \phi) H\dot{\phi}^2  - b_{,\phi} \dot{\phi}^3\right) \pi 
=0
\,,\nonumber\\
\label{offdiag}
&& \big(1+G_4(\phi)\big)  \left(\alpha - \psi  -\dot{\sigma} - H\sigma \right)  + G_{4,\phi} \left(\pi - \dot{\phi}\sigma \right)=0
\\
\label{diag}
& & \left( (1+G_4(\phi)) H + \frac12 G_{4,\phi}\dot{\phi} - \frac12 b( \phi) \dot{\phi}^3 \right) \dot{\alpha} 
\\
&+&  \left( (1+G_4(\phi)) \left(2 \dot{H} + 3 H^2 \right) + \Big( \frac12 G_{2,X} + G_{4,\phi\phi}\Big) \dot{\phi}^2  -  b_{,\phi} \dot{\phi}^4 + G_{4,\phi}\left(\ddot{\phi} + 2H\dot{\phi} \right)
- 2b(\phi)  \ddot{\phi} \dot{\phi}^2
\right)\alpha
\nonumber\\
&+& 
(1+G_4(\phi))\ddot{\psi}  + \left(3(1+G_4(\phi))H + G_{4,\phi}\dot{\phi}\right) \dot{\psi} - \frac12 \left(G_{4,\phi} - b( \phi) \dot{\phi}^2\right) (\ddot{\pi} + 3H\dot{\pi}) 
\nonumber\\
&-&  \frac12 \left( \left( G_{2,X} + 2G_{4,\phi\phi} \right) \dot{\phi} - G_{4,\phi}H - 2 b_{,\phi}\dot{\phi}^3 - b( \phi) \left( 2 \ddot{\phi} -3 H \dot{\phi} \right) \dot{\phi} \right) \dot{\pi}
\nonumber\\
&-&  \frac12 \left(G_{2,\phi} - b_{,\phi} \dot{\phi}^2 \ddot{\phi} - \frac{1}{2} b_{,\phi\phi} \dot{\phi}^4 + G_{4,\phi\phi\phi}\dot{\phi}^2 + G_{4,\phi\phi}\left(\ddot{\phi} + 2 H\dot{\phi}\right) 
+G_{4,\phi}(2\dot{H} + 3H^2) \right)\pi
 = 0 \,.
 \nonumber
\end{eqnarray} 
Here, Eq.~\eqref{0-0} is the linearized Hamiltonian constraint; Eq.~\eqref{t-i} is the linearized momentum constraint;  Eq.~\eqref{offdiag} is the linearized anisotropy constraint; 
and we used Eq.~\eqref{offdiag} to eliminate the $k$-dependence in the diagonal part of the linearized pressure equation~\eqref{diag}.

The equations significantly simplify if we substitute the field perturbation $\pi$ by $\delta u=-\pi/\dot{\phi}$. Note that working with $\delta u$, we constrain our analysis to $\dot{\phi} \neq 0$. But we can do this without loss of generality since in the vicinity of $\gamma$-crossing $\dot{\phi}\neq 0$.
\begin{eqnarray}
\label{N-0-0-u}
& & 
\left(3 \big(1 + G_4(\phi)\big) H^2 -\frac{1}{2}G_{2,X} \dot{\phi}^2 -\frac{1}{2} \big(G_{2,X X} - 2b_{,\phi}  \big) \dot{\phi}^4 -6 b( \phi) H \dot{\phi}^3  + 3 HG_{4,\phi}\dot{\phi}  \right) \alpha
\qquad\\
&+&  \left( \big(1 + G_4(\phi)\big)H - \frac12 b( \phi) \dot{\phi}^3 + \frac12 G_{4,\phi}\dot{\phi}   \right) \left(3\dot{\psi} +\frac{k^2}{a^2}\sigma \right)
+ \big(1+ G_4(\phi)\big) \frac{k^2}{a^2}  \psi
\nonumber\\
&-&   \frac{1}{2} \left( G_{2,X} \dot{\phi}^2 + \big(G_{2,X X} - 2b_{,\phi}  \big)  \dot{\phi}^4 + 9 b( \phi) H\dot{\phi}^3 - 3 G_{4,\phi}\dot{\phi}H \right)  \delta\dot{u} 
+ \frac{1}{2}\left(G_{4,\phi}\dot{\phi} - b( \phi)\dot{\phi}^3 \right) \frac{ k^2}{a^2}  \delta u 
\nonumber\\
&-&   \frac{1}{2} \left( G_{2,X} \dot{\phi} + \big(G_{2,X X} - 2b_{,\phi}  \big)  \dot{\phi}^3 + 9 b( \phi) H\dot{\phi}^2 - 3 G_{4,\phi}H \right)\ddot{\phi} \delta u
\nonumber\\
& + &  \frac{1}{2} \left( G_{2, \phi} - G_{2, X \phi} \dot{\phi}^2 - 3 b_{,\phi} H \dot{\phi}^3  + \frac{1}{2} b_{,\phi\phi} \dot{\phi}^4 + 3 H G_{4,\phi\phi}\dot{\phi} + 3 H^2 G_{4,\phi}  \right)\dot{\phi} \, \delta u 
= 0
\,,\nonumber\\
\nonumber\\
\label{N-t-i-u} 
& &  \left(  \big(1 + G_4(\phi)\big)H+ \frac12 G_{4,\phi} \dot{\phi} - \frac12 b( \phi)\dot{\phi}^3 \right) \alpha + \big(1 + G_4(\phi)\big) \dot{\psi} 
+ \frac12 \left(G_{4,\phi}\dot{\phi} - b( \phi) \dot{\phi}^3 \right) \delta \dot{u} 
\\
&+& \frac{1}{2} \left( \left(G_{4,\phi} - b( \phi) \dot{\phi}^2 \right) \ddot{\phi} + \left(G_{2,X} +G_{4,\phi\phi}\right) \dot{\phi}^2 - G_{4,\phi}\dot{\phi} H + 3b( \phi) H\dot{\phi}^3  - b_{,\phi} \dot{\phi}^4\right) \delta u
=0
\,,\nonumber\\
\nonumber\\
\label{N-offdiag-u}
&& \big(1+G_4(\phi)\big)  \left(\alpha - \psi  -\dot{\sigma} - H\sigma \right)  = G_{4,\phi}\dot{\phi}\, \left(\delta u + \sigma \right)
\,,
\\
\nonumber\\
\label{N-diag-u}
& & \left( (1+G_4(\phi)) H + \frac12 G_{4,\phi}\dot{\phi} - \frac12 b( \phi) \dot{\phi}^3 \right) \dot{\alpha} 
\\
&+&  \left( (1+G_4(\phi)) \left(2 \dot{H} + 3 H^2 \right) + \Big( \frac12 G_{2,X} + G_{4,\phi\phi}\Big) \dot{\phi}^2  -  b_{,\phi} \dot{\phi}^4 + G_{4,\phi}\left(\ddot{\phi} + 2H\dot{\phi} \right)
- 2b(\phi)  \ddot{\phi} \dot{\phi}^2
\right)\alpha
\nonumber\\
&+& 
(1+G_4(\phi))\ddot{\psi}  + \left(3(1+G_4(\phi))H + G_{4,\phi}\dot{\phi}\right) \dot{\psi} + \frac12 \left(G_{4,\phi}\dot{\phi} - b( \phi) \dot{\phi}^3\right) \delta\ddot{u}  
\nonumber\\
&+&   \left( \left( \frac12 G_{2,X} + G_{4,\phi\phi} \right) \dot{\phi}^2 + G_{4,\phi} \ddot{\phi} + G_{4,\phi}\dot{\phi}H -  b_{,\phi}\dot{\phi}^4 -  2b( \phi) \ddot{\phi}\dot{\phi}^2 \right) \delta\dot{u} 
\nonumber\\
&+&  \frac12 \left(G_{4,\phi} - b( \phi) \dot{\phi}^2\right) \dddot{\phi}\delta u +\frac12 \left( \left( G_{2,X} + 3G_{4,\phi\phi} \right) \dot{\phi} +  2G_{4,\phi}H - 2b_{,\phi}\dot{\phi}^3 -  2b( \phi) \ddot{\phi}\dot{\phi} \right) \ddot{\phi}\delta u 
\nonumber\\
&+&  \frac12 \left(G_{2,\phi} - b_{,\phi} \dot{\phi}^2 \ddot{\phi} - \frac{1}{2} b_{,\phi\phi} \dot{\phi}^4 + G_{4,\phi\phi\phi}\dot{\phi}^2 + 2G_{4,\phi\phi} \dot{\phi}H
+G_{4,\phi}(2\dot{H} + 3H^2) \right)\dot{\phi}\delta u
 = 0 \,.
 \nonumber
\end{eqnarray}

Note that the coefficient of $\ddot{\phi}\delta u$ in the pressure equation (second to last term in Eq.~\eqref{N-diag-u}) includes a triple time derivative $\dddot{\phi}$. This can be eliminated by rewriting the coefficient of $\ddot{\phi}\delta u$ term in Eq.~\eqref{N-diag-u} using the second Friedman equation and its time derivative:
\begin{eqnarray}
&& \left(G_{4,\phi} - b( \phi) \dot{\phi}^2\right) \dddot{\phi} + \left( \left(G_{2,X}  + 3G_{4,\phi\phi}\right)\dot{\phi} + 2HG_{4,\phi}  - 2b_{,\phi} \dot{\phi}^3  - 2b( \phi)\dot{\phi} \ddot{\phi} \right) \ddot{\phi}     =
 \\
&=& -2A_h\ddot{H} - 2\dot{A}_h\dot{H}   + 3H\left( G_{4,\phi}\ddot{\phi}  -b(\phi)\dot{\phi}^2\ddot{\phi} - b_{,\phi}\dot{\phi}^4 \right) + \left(b_{,\phi\phi} \dot{\phi}^4 - G_{2,X\phi}\dot{\phi}^2 - G_{4,\phi\phi\phi}\dot{\phi}^2 \right)\dot{\phi} 
\nonumber\\
&-&   \left(G_{2,X}  + G_{2,XX}\dot{\phi}^2 - 3b_{,\phi} \dot{\phi}^2 + 6Hb(\phi)\dot{\phi}\right)\dot{\phi}\ddot{\phi}   + G_{4,\phi\phi}\dot{\phi}^2H + \dot{H} \left(G_{4,\phi}\dot{\phi}- 3b(\phi)\dot{\phi}^3\right) 
\,.
\nonumber
\end{eqnarray}
The coefficient can be further simplified using the scalar-field equation~\eqref{FRW3}:
\begin{eqnarray}
&-& 2A_h\ddot{H} - 2\dot{A}_h\dot{H}   + 3H\left( G_{4,\phi\phi}\dot{\phi}^2 + G_{4,\phi}\ddot{\phi} -b(\phi)\dot{\phi}^2\ddot{\phi} - b_{,\phi}\dot{\phi}^4 +  G_{4,\phi}\dot{\phi}H  \right)
\\
&+&  \left(G_{2,\phi} - G_{2,X\phi}\dot{\phi}^2  +\frac{1}{2} b_{,\phi\phi} \dot{\phi}^4   - \left(G_{2,X}  + G_{2,XX}\dot{\phi}^2 - 2b_{,\phi} \dot{\phi}^2 + 6Hb(\phi)\dot{\phi}\right)\ddot{\phi}
 \right)\dot{\phi} 
\nonumber\\
&+& 3\dot{H} \left(G_{4,\phi} - b(\phi)\dot{\phi}^2\right)\dot{\phi} = -2\left(A_h\ddot{H} + \dot{A}_h\dot{H}   + 3A_hH\dot{H}\right)\,.
\nonumber 
\end{eqnarray}
Then Eq.~\eqref{N-diag-u} can be replaced by 
\begin{eqnarray}
& & \left( (1+G_4(\phi)) H + \frac12 G_{4,\phi}\dot{\phi} - \frac12 b( \phi) \dot{\phi}^3 \right) \dot{\alpha} 
\\
&+&  \left( (1+G_4(\phi)) \left(2 \dot{H} + 3 H^2 \right) + \Big( \frac12 G_{2,X} + G_{4,\phi\phi}\Big) \dot{\phi}^2  -  b_{,\phi} \dot{\phi}^4 + G_{4,\phi}\left(\ddot{\phi} + 2H\dot{\phi} \right)
- 2b(\phi)  \ddot{\phi} \dot{\phi}^2
\right)\alpha
\nonumber\\
&+& 
(1+G_4(\phi))\ddot{\psi}  + \left(3(1+G_4(\phi))H + G_{4,\phi}\dot{\phi}\right) \dot{\psi} + \frac12 \left(G_{4,\phi}\dot{\phi} - b( \phi) \dot{\phi}^3\right) \delta\ddot{u}  
\nonumber\\
&+&   \left( \left( \frac12 G_{2,X} + G_{4,\phi\phi} \right) \dot{\phi}^2 + G_{4,\phi} \ddot{\phi} + G_{4,\phi}\dot{\phi}H -  b_{,\phi}\dot{\phi}^4 -  2b( \phi) \ddot{\phi}\dot{\phi}^2 \right) \delta\dot{u} 
\nonumber\\
&-&  \left( A_h\ddot{H} + \dot{A}_h\dot{H}   + 3A_hH\dot{H} \right) \delta u
 = 0 \,.
 \nonumber
\end{eqnarray}

\section{Conformally-coupled ${\cal L}_4$ Horndeski in harmonic gauge}
\label{app-harmonicEqs}

Here, we derive the linearized Einstein equations in the harmonic gauge corresponding to the conformally-coupled ${\cal L}_4$ Horndeski action specified in Eq.~\eqref{G-i} 
and show that this gauge is well-behaved in the vicinity of and at the $\gamma$-crossing point ($\gamma=0$). This confirms our results obtained in the Newtonian gauge. 

The harmonic gauge \cite{Garfinkle:2001ni,Pretorius:2004jg} is defined as the foliation where the corresponding coordinates obey the condition
\begin{equation}
\label{harmonic-cooC}
\Box\, x^{\mu} = 0 
\,.
\end{equation}
A corollary is that, in harmonic gauge, the Christoffel symbols $\Gamma^{\rho}_{\mu\nu}$ must satisfy the constraint
\begin{equation}
\label{harmonic-connectionC}
g^{\mu\nu} \Gamma^{\rho}_{\mu\nu} = 0 
\,.
\end{equation}

Using the harmonic time coordinate $\tau$ that satisfies the gauge condition in Eq.~\eqref{harmonic-cooC}, the FRW line element takes the form
\begin{equation}
\label{harmonicFRW-lineE}
ds^2 = - a^6(\tau){\rm d}\tau^2 + a^2(\tau)\delta_{ij} {\rm d}x^i {\rm d} x^j 
\,,
\end{equation} 
and the homogeneous background equations read as
\begin{eqnarray}
\label{FRW1harm}
3 {\cal H} ^2  &=&  - a^{6} G_2(X,\phi) + G_{2,X}(X,\phi) \phi ^{\prime 2} - \frac{1}{2} b_{,\phi}(\phi) \frac{ \phi^{\prime 4}}{ a^6} + 3 {\cal H} b(\phi) \frac{ \phi ^{\prime 3}}{a^6} 
\nonumber\\
&-& 3G_{4,\phi}{\cal H}\phi' - 3G_4(\phi){\cal H}^2
\,, \\
\label{FRW2harm}
-2 {\cal H}'   &=&  2a ^{6} G_2(X,\phi) - G_{2,X}(X,\phi) \phi ^{\prime 2} - b(\phi)\frac{ \phi ^{\prime 2}}{a ^{6}}\phi''  + 2 G_{4,\phi}{\cal H}\phi' + G_{4,\phi\phi}\phi^{\prime 2} 
\nonumber\\
&+& G_{4,\phi}\phi'' + 2 G_4(\phi){\cal H}'
\,.
\end{eqnarray}
Here prime denotes the derivative with respect to harmonic time $\tau$. The harmonic time coordinate is related to the physical time coordinate $t$ via ${\rm d}t \equiv a^3(\tau){\rm d}\tau$
and the harmonic Hubble parameter ${\cal H}(\tau)$ is related to the physical Hubble parameter $H(t)$ via $H(t)={\cal H}(\tau)/a^3(\tau)$.

The evolution equation for the homogeneous scalar $\phi(t)$ is given by
\begin{eqnarray}
\label{FRW3harm-0}
&& \left(G_{2,X} + \left( G_{2,XX} - 2 b_{,\phi}\right) \frac{\phi^{\prime 2}}{a^6} + 6 b{\cal H} \frac{\phi'}{a^6}   \right) \phi''   
 - 3{\cal H} \frac{\phi^{\prime 3} }{a^6} \left( G_{2,XX} - 2b_{,\phi}\right)
+  3 b \frac{\phi^{\prime 2}}{a^6} \left( {\cal H}' - 6 {\cal H}^2 \right) =   
\nonumber\\ 
 &=& a^6 G_{2,\phi} - \left( G_{2,X\phi} - \frac{1}{2} b_{,\phi\phi} \frac{\phi^{\prime 2}}{a^6}\right) \phi ^{\prime 2} +3 G_{4,\phi}  \left({\cal H}' - {\cal H}^2 \right)
\,.
\end{eqnarray}

Further, using the harmonic time coordinate $\tau$, the linearized line element with scalar perturbations takes the form
\begin{equation}
\label{harmonic-lineE}
ds^2 = - a^6(\tau) \big(1+2\alpha \big){\rm d}\tau^2 + 2a^4(\tau) \partial_i \beta d\tau{\rm d} x^i + a^2(\tau)\left( \big( 1 - 2\psi \big)\delta_{ij} + 2\partial_i\partial_j\epsilon \right){\rm d} x^i {\rm d} x^j 
\,,
\end{equation} 
and the linearized harmonic gauge constraint as given in Eq.~\eqref{harmonic-connectionC} translates into two first-order partial differential equations 
\begin{eqnarray} 
\label{gauge1}
 && \alpha' + 3 \psi' + k ^{2} \left(\epsilon' - a ^{2} \beta \right) =0
 \,,\\  
 \label{gauge2}
&&\left( a ^{2} \beta \right)' + a ^{4} \left(\alpha - \psi + k ^{2} \varepsilon \right) =0\,.
\end{eqnarray}
Notably, the harmonic gauge conditions promote the lapse and shift perturbations to dynamical variables, and hence, the harmonic gauge is a differential gauge.

Rewriting Eqs.~(\ref{0-0}-\ref{diag}) in terms of the harmonic time coordinate $\tau$, the linearized Einstein equations for a single Fourier mode with wavenumber $k$ take the following form:
\begin{eqnarray}
\label{0-0-h}
& & 
\left(3 \big(1 + G_4(\phi)\big) \mathcal{H}^2 -\frac{1}{2}G_{2,X} \phi ^{\prime 2} -\frac{1}{2} \big(G_{2,X X} - 2b_{,\phi}  \big) \frac{ \phi^{\prime 4}}{a ^6} -6 b( \phi) \mathcal{H} \frac{ \phi^{\prime 3}}{a^6}   + 3 {\cal H}G_{4,\phi}\phi'  \right) \alpha 
\qquad\\
&+& \left( \big(1 + G_4(\phi)\big) {\cal H} - \frac12 b( \phi)  \frac{ \phi^{\prime 3}}{a^6} + \frac12 G_{4,\phi}\phi' \right) k^2(\varepsilon' - a^2\beta  )
\nonumber\\
&+&  \left( \big(1 + G_4(\phi)\big){\cal H} - \frac12 b( \phi) \frac{ \phi^{\prime 3}}{a^6} + \frac12 G_{4,\phi}\phi'   \right) 3\psi' 
+ \big(1+ G_4(\phi)\big) k^2 a^{4} \psi
\nonumber\\
&+&   \frac{1}{2} \left( G_{2,X} \phi' + \big(G_{2,X X} - 2b_{,\phi}  \big)  \frac{ \phi^{\prime 3}}{a^6} + 9 b( \phi) {\cal H} \frac{ \phi^{\prime 2}}{a^6} - 3 G_{4,\phi}{\cal H} \right) \pi' 
- \frac{1}{2}\left(G_{4,\phi} - b( \phi)\frac{ \phi^{\prime 2}}{a^6} \right)  k^2a^4 \pi 
\nonumber\\
&-&  \frac{1}{2} \left( a^6 G_{2, \phi} - G_{2, X \phi} \phi^{\prime 2} - 3 b_{,\phi} {\cal H} \frac{ \phi^{\prime 3}}{a^6}  + \frac{1}{2} b_{,\phi\phi} \frac{ \phi^{\prime 4}}{a^6} + 3 {\cal H} G_{4,\phi\phi}\phi' + 3 {\cal H} ^2 G_{4,\phi}  \right)\pi 
= 0
\,,\nonumber\\
\label{t-i-h} 
& &  \left(  \big(1 + G_4(\phi)\big) {\cal H} + \frac12 G_{4,\phi} \phi' - \frac12 b( \phi)\frac{\phi^{\prime 3}}{a^6} \right) \alpha + \big(1 + G_4(\phi)\big) \psi' \\
&-& \frac12 \left(G_{4,\phi} - b( \phi) \frac{\phi^{\prime 2}}{a^6} \right) \pi' 
- \frac{1}{2} \left( \left(G_{2,X} +G_{4,\phi\phi}\right) \phi' - G_{4,\phi} {\cal H} + 3b( \phi) {\cal H} \frac{ \phi^{\prime 2}}{a^6}  - b_{,\phi} \frac{ \phi^{\prime 3}}{a^6} \right) \pi 
=0
\,,\nonumber\\
\label{offdiag-h}
&& \big(1+G_4(\phi)\big)\Big( a^4 ( \alpha - \psi) - (\varepsilon' - a^2 \beta)'  \Big) + G_{4,\phi} \Big( a^4\pi- \phi'(\epsilon' -a^2\beta) \Big) =0
\\
\label{diag-h}
& & \left( (1+G_4(\phi)) {\cal H} + \frac12 G_{4,\phi}\phi' - \frac12 b( \phi) \frac{\phi^{\prime 3}}{a^6} \right) \alpha' 
+ (1+G_4(\phi)) \left(2 {\cal H}' - 3 {\cal H}^2 \right)\alpha
\\
&+& \left(  \left( \frac12 G_{2,X} + G_{4,\phi\phi}\right) \phi^{\prime 2}  -  b_{,\phi} \frac{\phi^{\prime 4}}{a^6} - 2b(\phi) \left( \phi'' - 3{\cal H} \phi' \right) \frac{\phi^{\prime 2}}{a^6} +G_{4,\phi}\left(\phi''-{\cal H}\phi'  \right) \right)\alpha
\nonumber\\
&+& (1+G_4(\phi))\psi'' + G_{4,\phi}\phi' \psi' 
-\frac12 \left(G_{4,\phi} - b( \phi) \frac{\phi^{\prime 2}}{a^6}\right) \pi'' 
\nonumber\\
&-& \frac12 \left( \left( G_{2,X} + 2G_{4,\phi\phi} \right) \phi' - G_{4,\phi}{\cal H} - 2 b_{,\phi}\frac{\phi^{\prime 3}}{a^6} - b( \phi) \left( 2 \phi'' -9 {\cal H} \phi' \right) \frac{ \phi'}{a^6} \right)\pi' 
\nonumber\\
&-& \frac12 \left(a^6 G_{2,\phi} - b_{,\phi} \frac{\phi^{ \prime 2}}{a^6} \left(\phi'' - 3 {\cal H} \phi'\right) - \frac{1}{2} b_{,\phi\phi} \frac{\phi^{ \prime 4}}{a^6}  + G_{4,\phi}(2{\cal H}' - 3{\cal H}^2)  \right)\pi
\nonumber\\
&-& \frac12 \left(G_{4,\phi\phi\phi}\phi^{\prime 2} + G_{4,\phi\phi}\left(\phi'' - {\cal H}\phi'\right)   \right)\pi =0
\nonumber
\,.
\end{eqnarray} 
As before, Eq.~\eqref{0-0-h} is the linearized Hamiltonian constraint; Eq.~\eqref{t-i-h} is the linearized momentum constraint;  Eq.~\eqref{offdiag-h} is the linearized anisotropy equation; and Eq.~\eqref{diag-h} is the linearized pressure equation.

In harmonic gauge, the linearized scalar-field equation is 
\begin{eqnarray}
&& \left( G_{2,X} \phi' + \left(G_{2,XX} - 2 b_{,\phi} \right) \frac{\phi^{\prime 3}}{a^6}  + 9 b( \phi) {\cal H} \frac{\phi ^{\prime 2}}{a^6} - 3G_{4,\phi}{\cal H} \right) \alpha' 
\\
&+ &  \left( 2 G_{2,X} \phi'' + G_{2,X \phi} \phi^{\prime 2} + G_{2,XX} \left( 5 \phi'' - 12 {\cal H} \phi' \right)  \frac{\phi^{\prime 2}}{a^6}
+ \left(G_{2,XX \phi}  - 2 b_{,\phi\phi}\right) \frac{\phi^{\prime 4}}{a^6}   \right) \alpha 
\nonumber
\\
& + &  \left(G_{2,XXX} \left( \phi'' - 3 {\cal H} \phi' \right)\frac{ \phi^{\prime 4}}{a^{12}} + 12b( \phi) \left( 2 {\cal H} \phi'' - 6 {\cal H}^{2} \phi'  + {\cal H}' \phi' \right)  \frac{\phi'}{a^6} 
- 8b_{,\phi} \left(  \phi'' - 3 {\cal H} \phi' \right) \frac{\phi^{\prime 2}}{a^6} \right) \alpha
\nonumber \\    
&+& 6G_{4,\phi} \left( {\cal H}^2 -  {\cal H}' \right)   \alpha
+ a^6 \left( G_{4,\phi} - b( \phi) \frac{\phi^{\prime 2}}{a^6}\right) \frac{k^2}{a^2} \alpha  
+ \left(  b( \phi) \frac{\phi^{\prime 2}}{a^6}   -  G_{4,\phi} \right) k^2 (\epsilon' -a^2\beta)'
\nonumber \\ 
&+ & \left( G_{2,X}\phi' + b( \phi) \left( 2 \phi'' -  3{\cal H} \phi' \right)\frac{ \phi'}{ a^6 }  - G_{4,\phi}{\cal H} \right) k^2 \left( \epsilon' - a^2 \beta \right) + 3\left( b(\phi)\frac{ \phi^{\prime 2} }{a^6} - G_{4,\phi} \right) \psi''
\nonumber \\
& - & 2a^6G_{4,\phi}  \frac{k^2}{a^2} \psi
+ 3 \left( G_{2,X} \phi'  + b( \phi)\left( 2\phi'' - 3 {\cal H} \phi' \right) \frac{ \phi'}{a^6 } - 3G_{4,\phi} {\cal H} \right) \psi'  
\nonumber
\\
& - &  \left(  G_{2,X} +  \left( G_{2,XX} - 2b_{,\phi} \right) \frac{\phi^{\prime 2}}{a^6} + 6 b( \phi) {\cal H} \frac{\phi'}{a^6}  \right) \pi'' -   \left( a^6 G_{2,X} + 2b( \phi) \left( \phi'' - {\cal H} \phi' \right) \right) \frac{k^2}{a^2} \pi
\nonumber \\ 
& - & 
\left( G_{2,X \phi} \phi' + 3G_{2,XX} \left( \phi'' - 3 {\cal H} \phi' \right) \frac{ \phi'}{ a^6}   + \frac{1}{a^6 } G_{2,XX \phi} \phi^{\prime 3}  + G_{2,XXX} \left( \phi'' - 3 {\cal H} \phi' \right) \frac{ \phi^{\prime 3} }{a^{12}}  \right) \pi' 
\nonumber \\ 
&- &
 \left( 6 \frac{b( \phi)}{a^6} \left( {\cal H} \phi'' - 6 {\cal H}^2 \phi' +  {\cal H}' \phi' \right) - 2b_{,\phi} (2 \phi'' - 9{\cal H} \phi')\frac{\phi' }{a^6}  - 2 b_{,\phi\phi} \frac{\phi^{\prime 3}}{a^6} \right) \pi' 
 \nonumber \\
&+ &
\left(  a^6 G_{2, \phi \phi} - G_{2,X \phi} \phi''  
- G_{2, X \phi \phi} \phi ^{ \prime 2} - \left( G_{2,XX \phi} - 2b_{,\phi\phi} \right) \left( \phi'' - 3 {\cal H} \phi' \right) \frac{ \phi^{ \prime 2}}{a^6 } \right) \pi
\nonumber\\
&- &
 \left( 3b_{,\phi} \left(  2{\cal H} \phi'' - 6 {\cal H}^2 \phi' + 3{\cal H}' \phi' \right) \frac{\phi'}{a^6}
 - \frac{1}{2} b_{,\phi\phi\phi} \frac{\phi^{ \prime 4}}{a^6} + 3 a^6 G_{4,\phi\phi} \left(  {\cal H}^2 -{\cal H}'\right) \right) \pi
=0  \,.\nonumber
\end{eqnarray}

Again, only three of the five field equations are independent. But,
together with the gauge conditions in Eqs.~\eqref{gauge1}~and~\eqref{gauge2}, the linearized Einstein equations form a complete set of five independent dynamical equations that fully determine the evolution of scalar perturbations. 
 
 As above, the linearized equations further simplify if we perform the variable change 
$
 \pi=-\phi'\delta u.
$
(Again, we can do this without loss of generality since in the vicinity of $\gamma$-crossing $\dot{\phi}\neq 0$.)
The following five equations form a closed system for the five scalar variables $\alpha, \beta, \psi, \epsilon$, and $\delta u$:
 \begin{eqnarray}
\label{0-0-u-harm}
&-& \Big( {\cal R}_K + 3 {\cal A}_h{\cal H}^2 - 6 {\cal H}  {\cal G}  \Big) \alpha + {\cal G} k^2(\varepsilon' - a^2\beta  )
+ 3 {\cal G}\psi' + {\cal A}_h k^2 a^{4} \psi  \qquad\\
&-&  \Big( {\cal R}_K  +  3{\cal H} \big({\cal A}_h{\cal H} - {\cal G}\big) \Big) \delta u' 
- \Big({\cal A}_h(\tau){\cal H}(\tau)-{\cal G}(\tau) \Big) k^2a^4 \delta u
- 3 {\cal G}(\tau) \left(  {\cal H}'  - 3  {\cal H}^2   \right) \delta u
= 0
\,,\nonumber\\
\label{t-i-u-harm} 
& &   {\cal A}_h \psi' -  \Big({\cal A}_h{\cal H} - {\cal G} \Big) \delta u' 
= -{\cal G} \alpha + {\cal A}_h \left(  {\cal H}'  - 3  {\cal H}^2   \right) \delta u 
\,,\\
\label{offdiag-u-harm}
&& {\cal A}_h(\tau)\Big( a^4 ( \alpha - \psi) - (\varepsilon' - a^2 \beta)'  \Big) = {\cal A}_h'(\tau)\Big( a^4\delta u + (\epsilon' -a^2\beta) \Big) \,,
\\
 && \alpha' + 3 \psi' + k ^{2} \left(\epsilon' - a ^{2} \beta \right) =0
 \,,\\  
&&\left( a ^{2} \beta \right)' + a ^{4} \left(\alpha - \psi + k ^{2} \varepsilon \right) =0\,.
\end{eqnarray}
Here, we introduced the harmonic coefficient functions
\begin{eqnarray}
{\cal A}_h(\tau)&=& 1+G_4(\phi) \,,
\\
{\cal G}(\tau) &=& \big(1 + G_4(\phi)\big){\cal H}(\tau) - \frac12 b( \phi) \frac{ \phi^{\prime 3}(\tau)}{a^6(\tau)} + \frac12 G_{4,\phi}\phi'(\tau) \,,
\\
{\cal R}_K(\tau)&=& \frac{1}{2}G_{2,X} \phi^{\prime 2}(\tau) + \frac{1}{2} \big(G_{2,X X} - 2b_{,\phi}  \big) \frac{ \phi^{\prime 4}(\tau)}{a^6(\tau)} + 3 b( \phi) {\cal H} (\tau) \frac{ \phi^{\prime 3}(\tau)}{a^6(\tau)}
\,,
\end{eqnarray}
where ${\cal A}_h(\tau) = A_h(t); {\cal G}(\tau) = a^3(t) \gamma(t)$; and ${\cal R}_K(\tau) = a^6(t) \rho_K(t)$. Obviously, $\gamma=0$ iff ${\cal G}=0$ and, hence, we call ${\cal G}=0$ `$\gamma$-crossing' as well.

To show that the system is non-singular in the vicinity of $\gamma$-crossing, we write it as a matrix equation
\begin{equation}
{\cal P}(\tau) y'(\tau, {\bf x}) = {\cal Q}(\tau)y(\tau, {\bf x}) \,,
\end{equation}
where $y^T = (\alpha \;\; a^2\beta \;\; \psi \;\; \epsilon \;\; \Xi \;\; \delta u)$ and the kinetic matrix is given by
\begin{equation}
{\cal P}(\tau)  = \left(\begin{array}{cccccc}
0 & 0 & 3{\cal G} & 0 & 0 & -( {\cal R}_K  +  3{\cal H} \big({\cal A}_h{\cal H} - {\cal G}\big) ) \\
0 & 0 & {\cal A}_h & 0 & 0 & -({\cal A}_h{\cal H} - {\cal G}) \\
0 & {\cal A}_h & 0 & 0 & -{\cal A}_h & 0\\
0 & 0 & 0 & 1 & 0 & 0\\
1 & 0 & 3 & 0 & 0 & 0 \\
0 & 1 & 0 & 0 & 0 & 0 
\end{array}\right) \,.
\end{equation}
The auxiliary variable $\Xi\equiv \epsilon'$ is needed to reduce the order of derivatives and make the system linear in all perturbation variables. 
The determinant of the kinetic coefficient matrix 
\begin{equation}
\det {\cal P}(\tau) = {\cal A}_h\left({\cal A}_h{\cal R}_K +3 \big({\cal A}_h {\cal H} - {\cal G}\big)^2  \right) \equiv {\cal A}_h\, \det P(t)
\end{equation}
is non-zero iff the determinant of the kinetic matrix $P(t)$ corresponding to the Newtonian-gauge equations~(\ref{0-0-newt}-\ref{offdiag-u-newt}) is non-zero, provided ${\cal A}_h\neq0$. 
In particular, the system is non-singular in the vicinity of and at $\gamma$-crossing.
 
We note that the use of harmonic gauge for studying the stability of certain Horndeski theories was previously considered in Ref.~\cite{Battarra:2014tga}.  However, it is important pedagogically to point out some flaws in their reasoning and analysis.  

First, the authors state that the harmonic gauge is needed to address a coordinate singularity near ${\cal H}=0$ (the bounce point).  This is incorrect:  as demonstrated in this paper, the evolution at ${\cal H}=0$ is generically well-behaved in conventional algebraic gauges (unitary, spatially flat, Newtonian, and synchronous) in braided theories.  This is true for the special ${\cal L}_3$ theories analyzed in Ref.~\cite{Battarra:2014tga}, and, as shown in this paper, in general  ${\cal L}_3$ and conformally-coupled ${\cal L}_4$ theories.  The key point, first made in Ref.~\cite{Ijjas:2016tpn}, is that braiding produces a blow-up at $\gamma$-crossing ($\gamma=0$), which occurs at a different time than the bounce (${\cal H} =0$).  Notably, $\gamma$ crossing  can be moved arbitrarily far from ${\cal H} =0$ (as measured by  time or field value).  That is, the blow-up that is in reality a coordinate singularity (as was shown in the present paper) is not associated with the bounce but with a different phenomenon in Horndeski theories.  In particular, in the algebraic gauges, $\alpha \sim 1/\gamma$ and $\sigma\sim 1/\gamma^2$ at $H=0$. That means, both the lapse and the shear (or shift) perturbations are generically non-singular in braided theories.  

Second, in braided theories, the quantity labeled ${\cal R}$ in Ref.~\cite{Battarra:2014tga} is {\it not}, as the authors describe, the co-moving curvature perturbation, a pointed out in Ref.~\cite{Kobayashi:2010cm} and also emphasized here.  Hence, its behavior at $\gamma$-crossing does not directly inform us about the gravitational stability of the theory.

Third, the numerical solutions on which the authors base their conclusions (especially Fig.~5) are inconclusive because it is not clear how sensitive the results are to initial conditions.  (We note that the caption refers to results for  two different ``gauge conditions''  where each is specified by fixing four different gauge variables.  Of course, a gauge choice fixes only two degrees of freedom; this adds further confusion about what the numerical experiments test.)

Fourth, they are claiming all solutions to the second-order differential equation have ${\cal R}'=0$ at $\gamma$-crossing. The second-order equation is derived in unitary gauge. But as a second-order differential equation it must have a second solution. If they are throwing it out because this gives a singular contribution -- that is not justified. Rather, they have to go to a gauge that gives good behavior.

Finally, Fig.~6, which is meant to be the key to the authors' proof of the stability of ${\cal L}_3$ Horndeski  bounces, is confusing. The caption suggests the six panels represent calculations in two different {\it gauges}.  The fact that ${\cal R}$ is the same in the Panel~$f$ is supposed to be the proof that the stabile behavior is gauge-invariant.  But there are a number of concerns about this figure. First, it claims two gauges but the caption actually describes two initial conditions, as noted above. Indeed, the first five panels show two solutions consistent with having two different initial conditions. When it comes to Panel~$f$, though, it appears to be only one curve and it is not clear what it represents.  It is evident (assuming $\Phi(t)$ represents the scalar field) that combining $\psi$ in Panel~$6c$ and $\Phi$ in Panel~$6e$ would give two different values of ${\cal R}$. So it is not clear what Panel~$6f$  represents and, hence, what it is proved by it. If it really were overlapping results of unitary and harmonic gauge calculations, then the unitary result would have to show bad behavior at $\gamma$-crossing (as noted above), which is not observed.

\section{Approximating the coefficients of the $\Psi$-equation~\eqref{psi-final-eq} in the vicinity of $\gamma$-crossing}
\label{sec:approx}

The goal of this Appendix is to show that the behavior of $\Psi$ at $\gamma$-crossing can be well-approximated analytically and shown to be well-behaved for wave-numbers below the braiding scale as given in Eq.~\eqref{def-k_B}, $k \lesssim k_B \lesssim 1$. 
More exactly, we derive the coefficients of Eq.~\eqref{psi-final-eq} that we use to analytically approximate the solutions of the $\Psi$-equation~\eqref{psi-final-eq} in the vicinity of $t = t_{\gamma}$, where $t_{\gamma}$ is the time of $\gamma$-crossing ($ \gamma(t_{\gamma})=0$). 

To describe the background, we expand $H, A_h,$ and $\gamma$ around $\gamma$-crossing  as given in Eqs.~(\ref{taylor-Ah}-\ref{taylor-gamma}). 
In addition, to keep all physical scales well below the Planck scale, we require 
\begin{eqnarray}
\label{a>0}
|t_{\gamma}| & \gg &  1
\,,\\
\label{a<10-3}
|H| & \simeq& p/|t_{\gamma}| \lesssim 10^{-3}
\,,\\
|\gamma | & \simeq& \gamma_0 |\Delta t| \lesssim 10^{-3}
\,;
\end{eqnarray}
and we normalize $a(t)$ so that $a(t_{\gamma})=1$.
As above, $\Delta t = t - t_{\gamma}$ indicates the time since $\gamma$-crossing, {\it i.e.}, $\Delta t =0$  at $\gamma=0$.  
In this Appendix, the goal is to approximate the behavior of $\Psi$ for a small interval of time around $t_{\gamma}$ to show that the solution is well-behaved at gamma-crossing; for this purpose, we can consider
\begin{equation}
\label{delta-t-constraint}
\left| \frac{ \Delta t}{t_{\gamma}} \right|  \ll 1.
\end{equation}
Note in particular that, in a bouncing scenario where $\gamma$-crossing occurs well before (or after) the bounce ($H=0$), our approximation for the background in Eqs.~(\ref{taylor-Ah}-\ref{taylor-gamma}) and \eqref{delta-t-constraint} applies near $\gamma$-crossing but does not extend all the the way to the bounce where $|\Delta t/t_{\gamma}| =1$. (Our earlier analysis in Ref.~\cite{Ijjas:2016vtq} already showed that the theory is perturbatively stable during the entire NEC-violating bounce stage and everywhere else other than in the vicinity of $\gamma=0$; so in this paper we only need to focus on $\gamma$-crossing.)

First, we show that for $k\lesssim k_B$ both $\det(P)$ and the denominator of the friction-term $\propto \dot{\Psi}$ and mass-term $\propto \Psi$  coefficients, 
\begin{equation}
\label{def-d}
d(t, k) = \det(P)\left(-\dot{H} + \frac{\dot{A}_h}{A_h} H \right)  +  \Big( A_hH - \gamma \Big)^2\frac{ k^2}{a^2}
\,,
\end{equation}
are nearly constant and the total coefficient of the friction term $F(t, k)$ in Eq.~\eqref{psi-final-eq} is nearly independent of $k$.
 Obviously, for arbitrary parameters $A_0, A_1, t_{\gamma}$, and $p$, the sign of $d(t, k)$ is not fixed because the term $-\dot{H} + (\dot{A}_h/{A_h} )H$ can be positive or negative. For simplicity of the analysis, in particular to avoid any artificial divergence of the `friction term' $\propto \dot{\Psi}$ and `mass-term' $\propto \Psi$ coefficients in Eq.~\eqref{psi-final-eq}, we only consider cases where $d > 0$ for all $k$. Together with the no-ghost condition ($\det(P)>0$), this imposes the constraint  
\begin{equation}
\label{d>0}
-\dot{H} + \frac{\dot{A}_h}{A_h} H > 0 
\,,
\end{equation}
{\it i.e.}, 
\begin{equation}
\frac{A_1}{A_0} t_{\gamma}^2 < 3
\,.
\end{equation}
For example, we can assume $(A_1/{A_0}) t_{\gamma}^2 \sim {\cal O}(1) <3$. We stress, though,  that the positivity constraint $d>0$ is not necessary. We only require it because it makes the stability analysis more straightforward.

With the conditions in Eqs.~(\ref{a>0}-\ref{delta-t-constraint}) and \eqref{d>0}, the expression for the kinetic matrix determinant can now be approximated as follows.
\begin{eqnarray}
\det(P) &=&A_h\Big(\dot{\gamma} + \ddot{A}_h - 4V + 9H\gamma + 5\dot{A}_hH + 2A_h\dot{H} + 3A_hH^2\Big) + 3\Big(A_hH-\gamma\Big)^2
\quad \\
&\simeq& A_0 \left(- 4 V_0 + 2 A_1 + \gamma_0  - 2A_0\frac{p(1-3p)}{t_{\gamma}^2} - p(9\gamma_0 + 10A_1)\left(-\frac{\Delta t}{t_{\gamma}}\right)  \right) 
\nonumber\\
&\simeq&  -4A_0V_0 
\,.\nonumber
\end{eqnarray}
Here, we assume that, in the vicinity of $\gamma$-crossing, the potential is well-approximated by $V\simeq V_0 \sim {\cal O}(-10^{-3}) <0$ (where $V_0=$ constant). This would be the case, for example, when $\gamma$-crossing occurs near the end of an ekpyrotic smoothing phase. To simplify the calculation, we can also assume $-4V_0 \gg A_1, \gamma_0$. (Again, the calculation can be easily generalized to different constellations of parameter values.)

We simplify the expression for the denominator $d$ and the friction $F$ in a similar way:
\begin{eqnarray}
\label{psi-final-eq-denom}
d(\Delta t) 
& \simeq& \frac{p }{ t_{\gamma}^2}(-4A_0V_0) \left( 1  + 2  \left( 1 - \frac{A_1}{A_0}
  t_{\gamma}^2\right) \left(-\frac{\Delta t}{t_{\gamma} }\right) +  \left(3 - 2 \frac{A_1}{A_0} t_{\gamma}^2  \right)\left(-\frac{\Delta t}{t_{\gamma} }\right)^2  \right)
\\
& \simeq&  \frac{p }{ t_{\gamma}^2}(-4A_0V_0) 
\,,
\nonumber\\
F(\Delta t) 
 & \simeq & \frac{2 }{t_{\gamma}} \left( \left(  1 -  \frac{A_1}{A_0} t_{\gamma}^2\right) \left( 1 -  3 \frac{\Delta t}{  t_{\gamma} } \right)
 + \left(6 - 5 \frac{A_1}{A_0} t_{\gamma}^2 
   + 5 \left(\frac{A_1}{A_0}t_{\gamma}^2 \right)^2 \right)\left(\frac{\Delta t}{t_{\gamma} }\right)^2 
 \right)
 \,.
 \end{eqnarray}
The fact that the friction term $F(t, k)$ is nearly $k$-independent above the braiding scale reduces its effect to a mere change in the time coordinate and, hence, does not affect linear stability.

Finally, to simplify the term $\propto \Psi$ in Eq.~\eqref{psi-final-eq}, we use the following approximations:
\begin{align}
& AH-\gamma \simeq -\frac{A_0}{(-t_{\gamma})}\left( p + \frac{\gamma_0 t_{\gamma}^2}{A_0}\left(-\frac{\Delta t}{t_{\gamma}}\right)\right)
\,,\\
& \frac{{\rm d}}{{\rm d}t} (AH-\gamma)^2 \simeq 2 \frac{A_0}{(-t_{\gamma})}\left( p + \frac{\gamma_0 t_{\gamma}^2}{A_0}\left(-\frac{\Delta t}{t_{\gamma}}\right)\right)
\,,\\
&-\dot{H} + \frac{\dot{A}_h}{A_h} H \simeq \frac{p}{t_{\gamma}^2}\left(1 + 2 \frac{A_1}{A_0}t_{\gamma}^2\left( \frac{\Delta t}{t_{\gamma}} + \left(-\frac{\Delta t}{t_{\gamma} } \right)^2  
 \right) \right)
\,, \\
& \frac{{\rm d}}{{\rm d}t} \left( -\dot{H} + \frac{\dot{A}_h}{A_h} H \right) \simeq
2 \frac{p}{t_{\gamma}^3}\left(\frac{A_1}{A_0}t_{\gamma}^2\right)\left( 1 - \left(-\frac{\Delta t}{t_{\gamma} } \right)  \right) 
\,;
\end{align}
to simplify our analytic expressions, we choose $\gamma_0 t_{\gamma}^2/A_0 \simeq {\cal O}(1)$. As exact numerical solutions show, our conclusions apply to a much wider range of parameters.
Substituting these approximations into Eqs.~(\ref{m_0-final}-\ref{u_H-final}) and keeping only terms up to second order in $t/t_{\gamma}$ yields the coefficient expressions given in Eqs.~(\ref{m_0-ap}-\ref{u_H-ap}).

\bibliographystyle{plain} 
\bibliography{newtonian-L4}

\end{document}